\documentclass[fleqn,usenatbib,useAMS]{article}
\usepackage{graphicx}				
\usepackage[T1]{fontenc}
\usepackage[utf8]{inputenc}
\usepackage{authblk}								
\usepackage{amssymb}
\usepackage{amsmath}
\usepackage{graphicx}
\usepackage[colorinlistoftodos]{todonotes}
\usepackage{dsfont}
\usepackage{mathrsfs}
\usepackage[title]{appendix}
\usepackage{newtxtext,newtxmath}
\usepackage[T1]{fontenc}
\usepackage{ae,aecompl}
\usepackage{graphicx} 
\usepackage{grffile}  
\usepackage{amsmath}  
\usepackage{amssymb}  

\newcommand{\chiPT}{$\chi${\it PT}}

\newcommand{\SUtw}{{\it SU(2)}}

\newcommand{\chiPTtw}{\SUtw~\chiPT}

\newcommand{\chiNL}{$\chi$NL}
\newcommand{\SchiNL}{Static\chiNL}
\newcommand{\GeV}{\mathit{GeV}}
\newcommand{\stimes}{\,\!{\times}\!\,}
\newcommand{\half}{\frac{1}{2}}
\newcommand{\bra}{\Big< \chi NL \Big\vert}
\newcommand{\ket}{\Big\vert \chi NL \Big>}
\newcommand{\brao}{{\phantom{\Big\vert}\!}_0\!\bra}
\newcommand{\keto}{\Big\vert \chi NL \Big>{\phantom{\Big\vert}\!\!\!}_0}

\newcommand{\one}{\mathds{1}}

\newcommand{\tildemone}{m^N\mathds{1}}

\graphicspath{ {images_sxnl/} }

\DeclareMathOperator{\Tr}{Tr}
\DeclareMathOperator{\expon}{exp}

\title{ Nuclides as a liquid phase of
$SU(2)_L \times SU(2)_R$ chiral perturbation theory\\ I: emergence of pion-less $SU(2)\chi PT$ 
}
\author[B.W. Lynn and G.D. Starkman]{
Bryan W. Lynn$^{1,2,3}$\thanks{E-mail: bryan.lynn@cern.ch}
~and 
Glenn D. Starkman$^{1,4}$\thanks{E-mail: glenn.starkman@case.edu}
\\
$^1$Dept Physics/CERCA/ISO, CWRU, Cleveland, OH, 44106 USA\\
$^2$Dept Physics /Astronomy, University College London, London WC1E 6BT, UK\\
$^3$Physics Division, CERN, 1211 Geneva, Switzerland\\
$^4$Dept Astronomy, Case Western Reserve University, Cleveland, OH, 44106 USA\\
}
\date{Accepted XXX. Received YYY; in original form ZZZ}
\begin{document}
\label{firstpage}
\maketitle

\begin{abstract} 

The Standard Model of particle physics (SM), augmented with neutrino mixing, is either
the complete theory of interactions of known particles
at energies accessible to Nature on Earth,
or very nearly so.  
Candidate effective theories of nuclear structure 
must therefore reflect SM symmetries, especially  the chiral global $SU(2)_L\times SU(2)_R$ symmetry of two-massless-quark QCD. For 
ground-state nuclei,
$SU(2) \chi PT$ enables
perturbation/truncation
in inverse powers of $\Lambda_{\chi SB}\simeq 1 GeV$, with analytic operators renormalized to all loop orders.
We show that
\chiPTtw ~ of protons, neutrons and 3 Nambu-Goldstone boson (NGB) pions admits a semi-classical ``liquid" 
phase,
 with energy required to
increase or decrease the density of constituents. 

We show that ``Pion-less'' \chiPTtw ~  emerges in the chiral liquid:
far-infrared NGB pions decouple from ``Static Chiral Nucleon Liquids (\SchiNL),"
vastly simplifying the derivation
of saturated nuclear matter (the infinite liquid phase) 
and of finite microscopic liquid drops (ground-state nuclides).
\SchiNL s explain the power of pion-less \chiPTtw~ 
to capture experimental ground-state properties of certain nuclides,
tracing that (no-longer-mysterious) empirical success 
directly to the global symmetries of two-massless-quark QCD.
\SchiNL~
are 
made entirely of nucleons.  
They have even parity; total spin zero;  
even proton number $Z$, and neutron number $N$;
and are arranged so 
local expectation values for spin and momenta vanish.

We derive the 
\SchiNL~ effective $SU(2)\chi PT$ Lagrangian, 
including all order $\Lambda_{\chi SB},\Lambda^0_{\chi SB}$ operators.
These include: all 
4-nucleon operators 
that survive Fierz rearrangement in the non-relativistic limit,
including operators that vanish for the non-relativistic \chiPTtw~  deuteron;
effective Lorentz-vector iso-vector neutral 
	``${\rho}$-exchange" operators
	crucial to $Z\neq N$ asymmetry effects. 
\SchiNL~ motivate nuclear matter, 
seen as non-topological solitons at zero internal and external pressure: the Nuclear Liquid Drop Model and 
Bethe-Weizs\"acker Semi-Empirical Mass Formula emerge in an
explicit Thomas-Fermi construction provided in the companion paper.
For chosen nuclides, nuclear Density Functional and Skyrme models 
are justified to order $\Lambda_{\chi SB}^0$. 
We conjecture that
 inclusion of 
	$\Lambda^{-1}_{\chi SB}$ and $\Lambda^{-2}_{\chi SB}$ 
operators will result in accurate "natural"  Skyrme,
No-Core-Shell, and 
ordinary 
neutron star 
models, 
with approximate liquid structure.

\end{abstract} 

\smallskip
\smallskip
\smallskip

\thanks{bryan.lynn@cern.ch}, 
 \thanks{glenn.starkman@case.edu}

\section{Introduction}
\label{Introduction}

\par 
In the Standard Model (SM) of particle physics, 
Quantum Chromodynamics (QCD) 
describes the strong interactions among quarks and gluons. 
At low energies, quarks and gluons are confined inside hadrons, 
concealing their degrees of freedom in such a way that 
we must employ an effective field theory (EFT) of hadrons. 
In doing so, we acknowledge as a starting point 
a still-mysterious experimental fact: 
Nature first makes hadrons 
and then assembles nuclei from them \cite{Ericson1988, Weinberg1990288, Ordonez1992459, vanKolck2002191}.

\par 
Since nuclei are made of hadrons, 
the fundamental challenge of nuclear physics is to
identify the correct EFT of hadrons
and use it to characterize all nuclear physics observations.
Many such EFTs have been considered \cite{Friar19963085, Burvenich2002044308, Finelli2004449, Rusnak1997495, Nikolaus19921757}.
Ultimately, the correct choice 
will both match the observations
and be derivable from the SM, i.e. QCD. 

Chiral perturbation theory (\chiPT) \cite{Weinberg19681568, Weinberg1978327, Coleman19692239, Callan19692247, Gasser1984142, Gasser1985465}  
is a low-energy perturbative approach to identifying 
the operators in the EFT that are allowed by the global symmetries of the SM.
It builds on the observation that the up ($m_{up}\simeq 6MeV$) 
	and down ($m_{down}\simeq 12 MeV$) quarks, 
	as well as the 3 pions ($\pi^\pm,\pi^0$, $m_{\pi}\simeq 140MeV$)--which are pseudo Nambu-Goldstone bosons--are all nearly massless compared to the other energy scales 
($\Lambda_{\chi SB} \simeq  1GeV$) in low-energy hadronic physics.

The effective-Lagrangian power counting \cite{Manohar1984189} of
$SU(2)_L\times SU(2)_R$ 
	chiral perturbation theory (\chiPTtw)   
incorporates all analytic higher-order quantum-loop corrections 
into tree-level amplitudes. 
The resultant perturbation expansion 
in the inverse of the chiral-symmetry-breaking scale 
$\Lambda^{-1}_{\chi SB}\simeq 1\GeV^{-1}$ 
renders \chiPTtw's strong-interaction predictions calculable in practice. 
Its low-energy dynamics of a proton-neutron nucleon doublet 
	and three pions as a pseudo-Nambu-Goldstone boson (pseudo-NGB) triplet
are our best understanding, together with lattice QCD, of the experimentally observed low-energy dynamics of QCD strong interactions. 
This understanding encompasses: 
pseudo-Nambu-Goldstone-boson (NGB) masses, 
soft-pion scattering, 
the applicability of $SU(2)_{L+R}\times SU(2)_{L-R}$ current algebra, 
the conserved vector current (CVC) 
and partially conserved axial-vector current (PCAC) hypothesis, 
semi-leptonic $\vec \pi$ decay, 
leptonic $\vec \pi$ decay, 
semi-leptonic nucleon decay, 
second class currents,
nucleon axial-vector couplings, 
the Goldberger-Treiman relation, 
nuclear beta decay (e.g. $^{14}O\rightarrow~ ^{14}\!N~e^+~\nu_e$),
precise measurement of Cabbibo angle,
{\it{et cetera}}.

\chiPTtw's effective-field-theoretic predictive power 
\cite{Weinberg19681568, Coleman19692239, Callan19692247, Gasser1984142, Gasser1985465, Manohar1984189, Georgi1984, Borasoy199885, Jenkins1991113}
derives from its ability to control its analytic quantum loops 
by power counting in $\frac{1}{\Lambda{\chi SB}}$, 
thus maintaining a well-ordered low-energy perturbation expansion 
that can be truncated. 
This predictive power stands in stark contrast 
with theories of strong interactions 
that lose their field-theoretic predictive power. 
These include any model of light or heavy nuclei 
not demonstrably derivable from the Standard Model \cite{deShalit1974}, 
such as
theories of quark bags 
and other confinement models of hadronic structure \cite{Farhi19842379, Farhi19852452}  
strange quark matter (and strange quark stars) 
\cite{Witten1984272, Chin19791292, Alcock1986261, Alcock19862088, Alcock19851273} 
and multi-Skyrmions in chiral pseudo-Goldstone symmetry \cite{Nikolaev1992101, Nikolaeva19921149, Jackson1989523}.

In contrast, 
QCD lattice-gauge-theory calculations of quarks and gluons 
\cite{Narayanan2005120, Muroya2003615, Richards2000} 
control their quantum loops, 
and we may hope that the detailed properties of the deuteron, 
the 
alpha particle,
and maybe even heavy nuclei, 
may someday be directly calculated in lattice QCD. 

Triumphant in claiming a role in nuclear physics,
\chiPTtw~
of dynamical nucleons and pions 
has been demonstrated \cite{Weinberg1965672, vanKolck19942932}  
to explain, to high accuracy, the detailed structure of the deuteron. 

B.W. Lynn \cite{Lynn1993281} first introduced the idea that $SU(2)\chi PT$ could also admit a liquid phase:
``It is legitimate to inquire whether the effective (power-counting) Lagrangian (\ref{LchiPTfull}) ... contains a liquid phase. An `$SU(2)_L\times SU(2)_R$ chiral liquid' is defined as a statistically significant number of baryons interacting via chiral operators ... with an almost constant (saturated) density ... (which) can survive as localized `(liquid) drops' at zero external pressure".  Lynn's 
Lagrangian included $SU(2)\chi PT$ terms of ${\mathcal O}(\Lambda_{\chi SB})$ and ${\mathcal O}(\Lambda_{\chi SB}^0)$ ignoring electro-magnetic breaking.
Anticipating the Static Chiral Nucleon Liquids studied here, he argued that, in the exact chiral limit, nucleons in the liquid phase interact with each-other only via the contact terms (\ref{BosonExchangeContactInteractions}). 
He did not derive pion-less $SU(2)\chi PT$. 
The study of chiral liquids in \cite{Lynn1993281} focused on those explicit chiral symmetry breaking terms whose origin lies entirely in non-zero light quark masses
$m_{up},m_{down}\neq 0$. 
(The $m=0, l=1, n=1$ contributions in (\ref{LchiPTfull})).
Here, we focus our study of chiral liquids instead on the  $n=0$ chiral limit, and prove the emergence of Pionless $SU(2)\chi PT$ in that chiral limit.

 \subsection{The fatal flaw in nuclear liquid drop models not based on \chiPTtw}

T.D.~Lee and G.C.~Wick \cite{Lee19742291} first identified 
	non-topological solitons 
	with the ground state of heavy-nuclei, as well as possible super-heavy nuclei,
thus making the crucial connection to nuclear liquids. 
Mathematically,  
such non-topological solitons emerge 
as a species of fermion Q-Ball \cite{Bahcall199067, Bahcall1989606, Selipsky1989430, Lynn1989465, Bahcall1998959, Bahcall1992}, or non-topological soliton \cite{Rosen1968999, Lee19742291, Lee1976254, Friedberg19762739, Friedberg197632, Coleman1985263, Friedberg19771694, Werle1977367, Morris1978337, Morris197887}.  
A practical goal would be to identify nuclear non-topological solitons 
with the ground state of ordinary even-even spin-zero spherically symmetric heavy nuclei, 
such as  $_{20}^{20}$Ca$_{40}$, $_{40}^{50}$Zr$_{90}$, 
	and $_{82}^{126}$Pb$ _{208}$. 

Nuclear non-topological solitons identified as nuclear liquids became popular  with the ingenious work of Chin and Walecka \cite{Chin197424} carried forward by \cite{Serot1979172}.
Nuclear ``Walecka models" \cite{Chin1977301, Serot19861, Serr197810} 
contain four dynamical particles: protons, neutrons, the Lorentz-scalar iso-scalar $\sigma$, 
and the Lorentz-vector iso-scalar $\omega_\mu$. 
\footnote{
	In practice, the $\omega_\mu$ is best treated 
	as a very heavy non-dynamical auxiliary field 
	and integrated out of the theory, 
	but, in order to be able to properly discuss 
	the renormalizability of the Walecka model, 
	we won't do so here. 
}
Nucleons are treated as 
	locally free-particles in Thomas-Fermi approximation. 
Finite-width nuclear surfaces are generated 
	by dynamical attractive $\sigma$-particle exchange, 
allowing them to exist at zero external pressure.
The empirical success of Walecka models is based on balancing $\sigma$ boson-exchange attraction
against $\omega_\mu$-boson-exchange repulsion. 
That that balance must be fine-tuned 
remains a famous mystery of the structure of the Walecka ground state.
In the absence of long-ranged electromagnetic forces, 
infinite symmetric  $(Z=N)$ nuclear matter, 
as well as finite microscopic ground-state $(Z=N)$ nuclides, 
appear as symmetric nuclear liquid drops. 

Both T.D.~Lee's and J.D.~Walecka's nuclear non-topological solitons 
are to be classified as ``liquids" because: 
\begin{itemize}
\item they have no crystalline or other ``solid" structure;
\item it costs energy to either increase or decrease the density 
	of the constituent nucleons compared to an optimum value;
\item they survive at zero external pressure, 
	e.g. in the absence of gravity, 
	so they are not a ``gas."
\end{itemize}


Despite their successes, 
there is a fatal flaw 
in all such current non-topological nuclear models, 
and in all nuclear models not based on \chiPTtw. 
To see this, examine the renormalizable
\footnote{
	Imagine $m_\omega^2$ arising from a spontaneously broken $U(1)$ gauge theory.
} 
tree-level Walecka Lagrangian:
\begin{eqnarray}
\label{WaleckaLagrangian}
L_{Walecka} &=& L_{Walecka}^{Nucleons} +L_{Walecka}^{\sigma} +L_{Walecka}^{\omega} \nonumber\\
L_{Walecka}^{Nucleons} &=& \overline{N} \Big[i\gamma^\mu \big(\partial_\mu +\omega_\mu \big) -m^N +g_\sigma \sigma \Big]N   \\
L^\sigma_{Walecka} &=&  \frac{1}{2}\big(\partial_\nu \sigma\big)^2 - V(\sigma ); \quad  
V(\sigma ) =  \frac{1}{2}m_\sigma^2 \sigma^2 + \frac{1}{4}\lambda_\sigma^2 \sigma^4 \nonumber \\
L^\omega_{Walecka} &=&  -\frac{1}{4}\Big[ \partial_\mu \omega_\nu -  \partial_\nu \omega_\mu\Big]\Big[ \partial^\mu \omega^\nu -  \partial^\nu \omega^\mu\Big] + \frac{1}{2}m_\omega^2 \omega_\mu \omega^\mu \nonumber
\end{eqnarray}
where $\frac{g_\sigma^2}{m_\sigma^2}=284.3GeV^{-2}$ 
	and $\frac{g^2_\omega}{m_\omega^2}=208.8 GeV^{-2}$ 
are fit to the experimentally inferred values 
of the number density $(k_{Fermi}\simeq1.42/fm)$ 
and saturated volume energy $(E_{binding}/nucleon \simeq 16MeV)$ 
of infinite symmetric $Z=N$ nuclear matter --
taken to be the interior of $^{82}_{126}Pb_{208}$ --
neglecting Coulomb and isospin effects.

The fatal flaw manifests when treating (\ref{WaleckaLagrangian}) 
as a quantum field theory beyond tree level.
Inclusion of 1-loop quantum corrections 
will strongly renormalize the values of 
$m^N$, $g_\sigma$, $m_\sigma^2$, $\lambda_\sigma$, $g_\omega$, 
	and $m_\omega^2$, 
and induce higher-order terms -- 
$\sim \sigma^6$,  $\sigma^{32}$, $\sigma^{784}$, ... -- 
with coefficients that depend on those parameters. 
We can dutifully re-fit (e.g. via Coleman-Weinberg) the 1-loop parameters to symmetric nuclear matter, 
including nuclear surface terms and compressibility, 
which now also depend on those new higher-power $\sigma$ interactions.
Next include 2-loop strong-interactions and re-fit.  
Because these are strong hadronic interactions, 
2-loop effects will be just as large as 1-loop effects, 
and cannot be truncated. 
Now include 3,4,5,..., 283 quantum loops 
(which are all required in any quantum field theory 
of strong hadronic interactions) and re-fit. 
Not only is such a program impossible in practice but, 
much worse, all the nuclear predictive power of the Walecka model 
has been completely lost!


This paper will cure those problems by strict compliance with the requirements of \chiPTtw~ effective field theory
of protons, neutrons and pions. 
The static chiral nucleon liquids (\SchiNL) studied below 
are true solutions to \chiPTtw. 
They include renormalized all-loop-orders analytic quantum corrections, are dependent on just a few 
experimentally measurable chiral coefficients, 
and restore theoretical predictive power over nuclides.

\section{The emergence of pion-less \SchiNL }

We recall  
the \chiPTtw~ Lagrangian
\footnote{
Important Infra-Red non-analytic terms in the pion sector are included in Appendix A
} 
to order $\Lambda_{\chi SB}$ and $(\Lambda_{\chi SB})^0$ 
in the chiral limit.


\begin{eqnarray}
\label{SymmetricLagrangian}
L^{Symmetric}_{\chi PT} 
	&=&   L^{\pi; Symmetric}_{\chi PT} 
		+ L^{N; Symmetric}_{\chi PT}  
		+ L^{4-N;Symmetric}_{\chi PT} \nonumber \\
L^{\pi; Symmetric}_{\chi PT} 
	&=& \frac{f^2_\pi}{4}\Tr \partial_\mu \Sigma \partial^\mu \Sigma^\dagger 
	+ L^{\pi;Symmetric}_{\chi PT;Non-Analytic}\\
L^{N; Symmetric}_{\chi PT} 
	&=& \overline{N} \Big(i\gamma^\mu (\partial_\mu + V_\mu) -  \tildemone \Big)N 
	- g_A\overline{N}\gamma^\mu\gamma^5{A_\mu N }  \nonumber\\
	&=& \overline{N} \Big(i\gamma^\mu \partial_\mu -  \tildemone \Big)N 
	+i{\vec J}^\mu \cdot{\vec V}_\mu 
	-g_A {\vec J}^{\mu,5}\cdot {\vec A}_\mu  \nonumber \\
L^{4-N;Symmetric}_{\chi PT} 
	&=& C_{\mathscr A} \frac{1}{2f^2_\pi}( \overline{N}\gamma^{\mathscr A} N)( \overline{N}\gamma_{\mathscr A} N) 
	+++\nonumber\,,
\end{eqnarray}
with fermion bi-linear currents
\begin{eqnarray}
{\vec J}^\mu &=&  \overline{N}\gamma^\mu {\vec t}N; \quad {\vec J}^{\mu, 5} =  \overline{N}\gamma^\mu \gamma^5 {\vec t}N \nonumber \\
V_\mu&=& {\vec t}\cdot {\vec V}_\mu; \quad {\vec V}_\mu = 2i\Big[\frac{\sin(\frac{{\pi}}{2f\pi})}{(\frac{\pi}{2f\pi})}\Big]^2{\vec \pi}\times \partial_\mu {\vec \pi} \nonumber \\
A_\mu&=& {\vec t}\cdot {\vec A}_\mu; \quad {\vec A}_\mu = -\frac{2}{{ \pi}^2}\Big[ {\vec \pi}\big({\vec \pi}\cdot \partial_\mu {\vec \pi}\big) 
+\frac{\cos(\frac{{\pi}}{2f\pi}) \sin(\frac{{\pi}}{2f\pi}) }{(\frac{\pi}{2f\pi})}{\vec \pi}\times\big(\partial_\mu{\vec \pi}\times {\vec \pi}\big)\Big] \nonumber 
\end{eqnarray}
where $\pi = \vert{\vec \pi}\vert =\sqrt{{\vec \pi}^2}$.

The parentheses in the four-nucleon Lagrangian 
indicate the order of $SU(2)$ index contraction,
and  $+++$  indicates that one should include 
all possible combinations of such contractions.
As usual,
  $\gamma^{\mathscr A} \equiv
  \left(1, \gamma^\mu,  i\sigma^{\mu\nu},  
  i\gamma^\mu \gamma^5,  \gamma^5\right)$,
for ${\mathscr A}=1,...,16$
(with 
$\sigma^{\mu\nu} \equiv \frac{1}{2}[\gamma^\mu,  \gamma^\nu ]$).
These are commonly referred to as 
scalar (S), vector (V), tensor (T), axial-vector (A), and pseudo-scalar (P)
respectively.
$C_{\mathscr A}$ are a set of chiral constants.


In the chiral limit, where $\vec \pi$s are massless, 
the presence of quantum nucleon sources 
could allow the massless NGB to build up, 
with tree-level interactions only, 
a non-linear quantum pion cloud. 
If we minimize the resultant action with respect to variations in the pion field, the equations of motion\footnote{
	This is a chiral limit  \chiPTtw~ analogue of QED where, in the presence of quantum lepton sources, a specific superposition of massless Infra-Red photons builds up into a classical electromagnetic field. Important examples are the ``exponentiation" of IR photons in $e^+e^-\rightarrow \mu^+\mu^-$ asymmetries, and $e^+e^-\rightarrow e^+e^-$ Bhabha scattering, at LEP1. Understanding the classical fields generated by initial-state and final-state soft photon radiation 
	\cite{Jadach1989201, Yennie1961379}   
	is crucial to dis-entangling high precision electro-weak loop effects, such as the experimentally confirmed precise Standard Model predictions for the 
	{ top-quark \cite{Lynn1984} and Higgs'\cite{Lynn1984, Djouadi1987265} masses}.
} 
capture the part of the quantum cloud 
that is to be characterized as a classical soft-pion field, 
thus giving us the pion ground-state 
(and content/configuration/structure) 
in the presence of the ground-state 
``Chiral Nucleon Liquid" \chiNL~ with fixed baryon number $A=Z+N$
\begin{eqnarray}
	0&=&\Big[\partial_\nu \frac{\partial}{\partial \big(\partial_\nu\pi^m\big)} 
		 -\frac{\partial}{\partial \pi^m} \Big] L_{\chi PT}^{Symmetric}  \\
	&=&\Big[\partial_\nu \frac{\partial}{\partial \big(\partial_\nu\pi^m\big)} 
	 	 -\frac{\partial}{\partial \pi^m} \Big] L_{\chi PT}^{\pi ;Symmetric}
	 	 \nonumber \\
	&+& i{\vec J}^\mu \cdot \Big[\partial_\nu \frac{\partial}{\partial \big(\partial_\nu\pi^m\big)} -\frac{\partial}{\partial \pi^m} \Big]{\vec V}_\mu 
	-g_A {\vec J}^{\mu,5}\cdot \Big[\partial_\nu \frac{\partial}{\partial \big(\partial_\nu\pi^m\big)} -\frac{\partial}{\partial \pi^m} \Big]{\vec A}_\mu \nonumber \\
	&-& 2  \partial_\mu {\vec J}^\mu \cdot 
		\left(\frac{\sin(\frac{{\pi}}{2f\pi})}{(\frac{\pi}{2f\pi})}\right)^2
		\left({\vec \pi}\times {\hat m}\right)\nonumber
	\\
	&+&\frac{2}{{ \pi}^2} g_A \partial_\mu{\vec J}^{\mu,5}\cdot \Big[ {\vec \pi}\big({\vec \pi}\cdot {\hat m}\big) 
	+\frac{\cos(\frac{{\pi}}{2f\pi}) \sin(\frac{{\pi}}{2f\pi}) }{(\frac{\pi}{2f\pi})}{\vec \pi}\times\big({\hat m}\times {\vec \pi}\big)\Big] 
	\nonumber
\end{eqnarray}

 We divide the classical pion field into 
 ``Infra-Red" and ``Non-IR" parts.
By definition, only ``IR" pions survive the internal projection operators 
associated with taking expectation values 
of the classical NGB $\vec \pi$s in the $\ket$ quantum state
\begin{eqnarray}
\label{IRPions}
	\bra  Function\big( \partial_\mu {\vec \pi}, {\vec \pi} \big)\ket
	\!\!\!\!\!\!\!\!\!\!\!\!\!\!\!\!\!\!\!\!\!\!\!\!\!\!\!\!\!\!\!\!\!&&  \\
	&=& \bra IR Part Of\Big[Function\big( \partial_\mu {\vec \pi}, {\vec \pi} \big)\Big]\ket \nonumber \\
	&\equiv& \Big\{ Function\big( \partial_\mu {\vec \pi}, {\vec \pi} \big) \Big\}_{IR} \nonumber \\
	0 &=& \bra Non-IR Part Of\Big[Function\big( \partial_\mu {\vec \pi}, {\vec \pi} \big)\Big] \ket \nonumber
\end{eqnarray}
 
The IR part does not change the \chiNL.
It could in principle be an important part of the \chiNL: 
	a $\vec \pi$ condensate,
	a giant resonance,
	a breathing mode,
	a time-dependent flashing-pion mode.
To ignore such classical IR $\vec \pi$s 
would therefore be an incorrect definition of \chiNL.
For finite \chiNL, 
it could be just a passing pion (of any frequency) 
which simply does not strike the \chiNL.

We call these ``IR pions" by keeping in mind a simple picture, 
where the $\vec \pi$ wavelength is ``long", 
i.e. longer than the scale within the \chiNL~  
	over which the local mean values of nucleon spin and momentum vanish.
Only ``IR" pions survive the internal projection operators associated with taking expectation values of the classical NGB $\vec \pi$s in the $\ket$ quantum state

 We now take expectation values of the $\vec \pi$ equations of motion. 
 In the presence of the quantum \chiNL~  source,
 the classical NGB $\vec \pi$ cloud obeys
 \begin{eqnarray}
 \label{EoMExpectation}
 0&=&\bra \Big[\partial_\nu \frac{\partial}{\partial \big(\partial_\nu\pi^m\big)} -\frac{\partial}{\partial \pi^m} \Big] L_{\chi PT}^{Symmetric} \ket \nonumber \\
&=&\Big\{ \Big[\partial_\nu \frac{\partial}{\partial \big(\partial_\nu\pi^m\big)} -\frac{\partial}{\partial \pi^m} \Big] L_{\chi PT}^{\pi ;Symmetric}\Big\}_{IR} \\
&+& i \bra {\vec J}^\mu \ket \cdot \Big\{\Big[\partial_\nu \frac{\partial}{\partial \big(\partial_\nu\pi^m\big)} -\frac{\partial}{\partial \pi^m} \Big]{\vec V}_\mu\Big\}_{IR} \nonumber\\
&-&g_A \bra {\vec J}^{\mu,5} \ket \cdot \Big\{\Big[\partial_\nu \frac{\partial}{\partial \big(\partial_\nu\pi^m\big)} -\frac{\partial}{\partial \pi^m} \Big]{\vec A}_\mu\Big\}_{IR} \nonumber \\
&-&  2 \bra \partial_\mu {\vec J}^\mu \ket \cdot \Big\{\left(\frac{\sin(\frac{{\pi}}{2f\pi})}{(\frac{\pi}{2f\pi})}\right)^2{\vec \pi}\times {\hat m} \Big\}_{IR}\nonumber \\
&+& \frac{2}{{ \pi}^2} g_A \bra \partial_\mu{\vec J}^{\mu,5} \ket \cdot \Big\{{\vec \pi}\big({\vec \pi}\cdot {\hat m}\big) 
+\frac{\cos(\frac{{\pi}}{2f\pi}) \sin(\frac{{\pi}}{2f\pi}) }{(\frac{\pi}{2f\pi})}{\vec \pi}\times\big({\hat m}\times {\vec \pi}\big)\Big\}_{IR}\nonumber
\end{eqnarray}


Examining the ground-state expectation values of the nucleon currents
and their divergences in \eqref{EoMExpectation}, we find that almost all of them vanish:
\begin{eqnarray}
\label{VanishingCurrents1}
\bra {J}_{\mu}^\pm \ket &=& 0\,,
	\quad \bra {J}_{\mu}^{\pm,5} \ket =0\,, \\
\bra \partial^\mu{J}_{\mu}^\pm \ket &=& 0\,,
	\quad \bra \partial^\mu{J}_{\mu}^{\pm,5} \ket =0\,,\nonumber
\end{eqnarray} 
because ${J}_{\mu}^\pm$ and ${J}_{\mu}^{\pm,5}$ change neutron and proton number.
Since the liquid ground state is homogeneous and isotropic, 
spatial components of vector currents vanish, in particular
\begin{eqnarray}
\label{VanishingCurrents3}
\bra {J}_{i}^3 \ket \simeq 0\
\end{eqnarray}
for Lorentz index $i=1,2,3$.
Because left-handed and right-handed protons and nucleons
are equally represented in the nuclear ground state,
\begin{eqnarray}
\label{VanishingCurrents4}
\bra {J}_{\mu}^{3,5} \ket \simeq 0\
\end{eqnarray}
for all $\mu$.
Current conservation {(see section 4)} enforces 
\begin{eqnarray}
\label{VanishingCurrents5}
\bra \partial^\mu{J}_{\mu}^{3} \ket  = 0\,,\quad
\bra \partial^\mu{J}_{\mu}^{3,5} \ket  = 0\,.
\end{eqnarray}
This leaves only a single non-vanishing current expectation value.
\begin{eqnarray}
 \label{VanishingCurrents}
  \bra {J}_{0}^3 \ket &\neq& 0 \,.
\end{eqnarray}

Equation \eqref{EoMExpectation},  
governing the classical pion cloud,
is thus enormously simplified
 \begin{eqnarray}
	0 &\simeq& \Big\{
	 \Big[\partial_\nu \frac{\partial}{\partial \big(\partial_\nu\pi^m\big)} - \frac{\partial}{\partial \pi^m} \Big] L_{\chi PT}^{\pi ;Symmetric}\Big\}_{IR} \\
	&&+ i \bra {J}^{3;0} \ket \Big\{\Big[\partial_\nu \frac{\partial}{\partial \big(\partial_\nu\pi^m\big)} -\frac{\partial}{\partial \pi^m} \Big]{V}_0^3 \Big\}_{IR}\nonumber
 \end{eqnarray}
with
 \begin{eqnarray}
 	 \label{SoftPionEnergy}
	&&\!\!\!\!\!\!\!\!\Big\{\Big[\partial_\nu \frac{\partial}{\partial \big(\partial_\nu\pi^m\big)} -\frac{\partial}{\partial \pi^m} \Big]{V}_0^3\Big\}_{IR}
	 =\\
	 &&\!\!\!\!\!\!\!\!\quad\Big\{
	2i\Big[\big(\partial_0{\vec \pi}\big)\times {\hat m}+{\vec \pi}\times {\hat m}\partial_0 -{\hat m}\times \big(\partial_0{\vec \pi}\big)  
	-{\vec \pi}\times \big(\partial_0{\vec \pi}\big)\frac{\partial}{\partial \pi^m}  \Big]^3 ~~
	\frac{\sin^2(\frac{{\pi}}{2f\pi})}
	{(\frac{\pi}{2f\pi})^2}
	 \Big\}_{IR}\nonumber\,.
\end{eqnarray}


A crucial observation is that (\ref{SoftPionEnergy}) 
is linear in $\partial_0{\vec \pi}$, 
i.e. in the energy of the classical NGB IR $\vec \pi$ field.
Expecting the nuclear ground state, 
and thus its  classical IR $\vec\pi$ field, 
to be static, we enforce 
\begin{equation}
\label{StaticPion}
\Big\{\partial_o{\vec \pi}  \Big\}_{IR} = 0\,.
\end{equation}
It now follows that 
\begin{eqnarray}
\label{PionDecouplinga}
 \Big\{\Big[\partial_\nu \frac{\partial}{\partial \big(\partial_\nu\pi^m\big)} -\frac{\partial}{\partial \pi^m} \Big]{V}_0^3 \Big\}_{IR} &=&0 \,.
\end{eqnarray}
independent of $\bra {J}^{3;0} \ket$. 
The IR pion equation of motion
\begin{eqnarray}
\label{PionDecouplingb}
  \Big\{
 \Big[\partial_\nu \frac{\partial}{\partial \big(\partial_\nu\pi^m\big)} - \frac{\partial}{\partial \pi^m} \Big] L_{\chi PT}^{\pi ;Symmetric}\Big\}_{IR} &=&0  \,
\end{eqnarray}
therefore has no nucleon source.
{
$L_{\chi PT}^{\pi ;Symmetric}$ in \eqref{PionDecouplingb} includes both its
analytic and non-analytic contributions 
(cf. appendix equation \eqref{LpiSymmetricnonAnalytic}).
The ground-state nucleons are not a source of any 
static IR NGB $\vec \pi$ classical field.}

The nuclear ground state in the chiral liquid is thus a static chiral nucleon liquid 
(\SchiNL), with no $\vec \pi$ condensate
\footnote{
	After explicit chiral symmetry breaking, 
	with non-zero $u,d$ quark and resultant pion masses, 
	and with Partially Conserved  Axial Currents (PCAC), 
	a static  S-wave $\vec \pi$ condensate 
	is a logical possibility \cite{Lynn1993281}, 
}
or time-dependent pion-flashing modes.  
We now write $\keto$ for the ground state 
to emphasize that it is static.


 We want to quantize the nucleons 
 in the background field of the static \chiNL, 
 and so consider the  expectation value of the nucleon equation of motion 
 in the chiral nucleon liquid ground state:
\begin{eqnarray}
\label{ChiralLiquidDiracEquation}
0&=& \brao \overline{N} \frac{\partial}{\partial \overline{N} }L^{Symmetric}_{\chi PT} \keto \\
&=& \brao \overline{N} \Big(i\gamma^\mu \partial_\mu -  \tildemone \Big)N \keto \nonumber \\
&+& i \brao{\vec J}^\mu\keto \cdot \Big\{ {\vec V}_\mu \Big\}_{IR} -g_A\brao {\vec J}^{\mu,5}\keto \cdot \Big\{ {\vec A}_\mu  \Big\}_{IR}\nonumber \\
&+& \frac{1}{f^2_\pi}\brao C_{\mathscr A}( \overline{N}\gamma^{\mathscr A} N)( \overline{N}\gamma_{\mathscr A} N) +++ \keto\nonumber\,.
\end{eqnarray}
Since
most of the nucleon $SU(2)_L\times SU(2)_R$ currents 
vanish in the \SchiNL,
and $\Big\{\partial_o{\vec \pi}  \Big\}_{IR} = 0$,
\begin{eqnarray}
\label{StaticChiralLiquidDiracEquation}
0 &\simeq& \brao \overline{N} \Big(i\gamma^\mu \partial_\mu -  \tildemone \Big)N \keto \\
&&+ \frac{1}{f^2_\pi}\brao C_{\mathscr A}( \overline{N}\gamma^{\mathscr A} N)( \overline{N}\gamma_{\mathscr A} N) +++ \keto\,. \nonumber
\end{eqnarray}


Equations (\ref{PionDecouplingb}) 
	and (\ref{StaticChiralLiquidDiracEquation}) 
show that, 
to order  $\Lambda_{\chi SB}$ and $\big(\Lambda_{\chi SB}\big)^0$,
\SchiNL~  are composed entirely of nucleons. 
That is also the basic premise of many empirical models 
	of the nuclear ground state: 
	Pion-less \chiPTtw, 
	Weizs\"acher's Semi-empirical Mass Formula, 
	the Nuclear Liquid Drop Model, 
	Nuclear Density Functional Models, 
	no-core Nuclear Shell Models, and Nuclear Skyrme Models. 
We have shown that that empirical nuclear premise 
can be (approximately) traced directly 
to the global $SU(2)_L\times SU(2)_R$ symmetries 
	of 2-massless-quark Quantum Chromodynamics, 
i.e. directly to the Standard Model of elementary particles.

The effective Lagrangian
derived from $SU(2)_L\times SU(2)_R \chi PT$ 
governing \SchiNL~  can now be written
\begin{eqnarray}
\label{NucleonLiquidLagrangian}
L_{Static\chi NL} &=& L^{Free Nucleons}_{Static\chi NL}  +L^{4-N}_{Static\chi NL} \\
L^{Free Nucleons}_{Static\chi NL} &=& \brao \overline{N} \Big(i\gamma^\mu \partial_\mu -  \tildemone \Big)N   \keto \nonumber \\
L^{4-N}_{Static \chi NL} &=& \brao \frac{1}{2f^2_\pi}C_{\mathscr A} ( \overline{N}\gamma^{\mathscr A} N)( \overline{N}\gamma_{\mathscr A} N) +++ \keto  \nonumber\,,
\end{eqnarray}

Pion-less \chiPTtw~ thus emerges inside  nuclear \SchiNL.
Within all-analytic-orders renormalized \chiPTtw,
infrared NGB pions effectively decouple from \SchiNL, 
vastly simplifying the derivation 
of the properties  of saturated nuclear matter 
	(the infinite liquid phase) 
and of finite microscopic liquid drops (the nuclides). 
\SchiNL~ thus explain 
the (previously puzzling) power of pion-less \chiPTtw~ 
to capture experimental ground-state facts 
of certain nuclides, 
by tracing that (no-longer-mysterious) empirical success 
directly to the global symmetries of two-massless-quark QCD.

It will be shown below that static \chiNL s 
satisfy all relevant $SU(2)_L\times SU(2)_R$ 
vector and axial-vector current-conservation equations 
in the liquid phase.\newline
\SchiNL~  are therefore solutions 
of the semi-classical liquid equations of motion; 
they are not just an ansatz.

\section{Semi-classical \SchiNL~   as the approximate ground state of certain nuclei}


To further elucidate the properties of the \SchiNL,
we must address the four-nucleon interactions.
This is best done by resolving $L_{Static \chi NL}^{4-N}$
into terms that are the products of two ground-state current expectation values --
usually described as boson-exchange contact interactions --
and terms that are the products of two transition matrix elements between
the ground state and an excited nuclear state,
\begin{eqnarray}
L_{Static \chi NL}^{4-N} 
&=& L_{Static \chi NL}^{4-N;BosonExchange} 
	+ L_{Static \chi NL}^{4-N;ExcitedNucleon} \,.
\end{eqnarray}


{\it A priori} there are 10 possible contact interactions representing
isosinglet and isotriplet  channels for each of five ``spatial''  current types:
scalar, vector, tensor, pseudo-scalar and axial-vector,
and so  {\it{10}} chiral coefficients parametrizing 4-nucleon contact terms:
$C^{T=0}_{S}$,$C^{T=1}_{S}$,$C^{T=0}_{V}$, $C^{T=1}_{V}$,$C^{T=0}_{T}$,$C^{T=1}_{T}$, $C^{T=0}_{P}$,$C^{T=1}_{P}$, $C^{T=0}_{A}$,and $C^{T=1}_{A}$.
\footnote{
In this Section, we will adopt the shorthand
$\Big<  \equiv \brao$ as well as $\Big> \equiv \keto$
}

The inclusion of exchange interactions 
induces the isospin ($T=1$) operators 
	to appear \cite{Lynn1993281}, 
and potentially greatly complicates 
	the effective chiral Lagrangian.
Fortunately, we are interested here 
	in the liquid limit of this Lagrangian.
Spinor-interchange contribution 
	are properly obtained by Fierz rearranging first, 
then imposing the properties of the semi-classical liquid (see Appendix B).
The appropriate \SchiNL~  Lagrangian, 
	and the resulting Dirac equation, 
	are consequently reasonably simple.
In fact, 
for the Static$\chi$NL,
the contact interactions can be represented by
\begin{eqnarray}
\label{BosonExchangeContactInteractions}
-L_{Static \chi NL}^{4-N;BosonExchange} &=&\frac{1}{2f^2_\pi} C^{S}_{200} \Big\{  \Big< {\overline N}N \Big> \Big< {\overline N}N \Big>  \Big\}\\
 &-& \frac{1}{4f^2_\pi} {\overline {C^{S}_{200}}}  \Big\{ \Big< {\overline N}N \Big> \Big< {\overline N}N \Big> 
+ 4\Big< {\overline N}t_3 N \Big> \Big< {\overline N}t_3 N \Big> 
 \Big\} \nonumber \\
&+&\frac{1}{2f^2_\pi} C^{V}_{200} \Big\{  \Big< {N^{\dagger}}N \Big> \Big< {N^{\dagger}}N \Big>  \Big\}\nonumber\\
 &-& \frac{1}{4f^2_\pi} {\overline {C^{V}_{200}}}  \Big\{ \Big< {N^{\dagger}}N \Big> \Big< {N^{\dagger}}N \Big> 
+ 4\Big< {N^{\dagger}}t_3 N \Big> \Big< {N^{\dagger}}t_3 N \Big> 
 \Big\}\nonumber\,,
\end{eqnarray}
with only {\it{4}} independent chiral coefficients:
\begin{eqnarray}
\label{NuclearCoefficients}
C^{S}_{200}&=&C^{T=0}_{S} \nonumber \\
-{\overline {C^{S}_{200}}} &=& \half \left[ \half C^{T=0}_{S}+ \frac{5}{4} C^{T=1}_{S}+3\Big( C^{T=0}_{T} +\half C^{T=1}_{T}\Big) +\half \Big( C^{T=0}_{P}+ \half C^{T=1}_{P} \Big) \right] \nonumber \\
C^{V}_{200}&=&C^{T=0}_{V}  \\
-{\overline {C^{V}_{200}}} &=& \half \left[ -C^{T=0}_{V}+C^{T=0}_{A} +\half C^{T=1}_{V}+\half C^{T=1}_{A}\right] \,.\nonumber
\end{eqnarray}
%
This is a vast improvement in the predictive power of the theory,
while still providing sufficient free parameters 
to balance vector repulsive forces against scalar attractive forces, 
when fitting (to order  $(\Lambda_{\chi SB})^0$) 
Non-topological Soliton, Density Functional and Skyrme nuclear models 
to the experimental structure of ground-state nuclei.

There is yet another simplification for a sufficiently large number of nucleons: simple Hartree analysis of (\ref{BosonExchangeContactInteractions}) is equivalent to far-more accurate Hartree-Fock analysis of the same Lagrangian without spinor-interchange terms.

More coefficients would be required to 
parametrize the excited-nucleon interactions:
\begin{eqnarray}
\label{ExcitedNucleon}
  -L_{Static \chi NL}^{4-N;ExcitedNucleon} 
  \!\!\!\!\!\!\!\!\!\!\!\!\!\!\!\!\!\!\!\!\!\!\!\!\!\!\!\!\!\!\!\!\!\!\!\!\!\!\!\!\!\!\!\!\!\!\!\!\!\!\!\!\!\!\!\!\!\!\!\!\!\!\!
  &&\\
  &=& \frac{1}{ 2 f_\pi^2}
  	 \sum_{\Psi \neq \vert\chi NL\rangle_{\!0}} \sum_{\mathscr A}
  	\bigg[ C^{T=0}_{\mathscr A} \brao  \overline{N_c^\alpha}\gamma^{{\mathscr A}\alpha\beta} N_c^\beta)
\big\vert \Psi \big>\big< \Psi \big\vert
( \overline{N_e^\lambda}\gamma_{\mathscr A}^{\lambda\sigma} N_e^\sigma) \keto  \nonumber \\
 &+&
 \sum_B C^{T=1}_{\mathscr A}\brao \frac{1}{4}( \overline{N_c^\alpha}\sigma^B_{cd}\gamma^{{\mathscr A}\alpha\beta} N_d^\beta)\big\vert \Psi \big>\big< \Psi \big\vert
( \overline{N_e^\lambda}\sigma^B_{ef}\gamma_{\mathscr A}^{\lambda\sigma} N_f^\sigma)\keto \bigg]\,.\nonumber
\end{eqnarray}
%
%
However, 
excited-nuclear contributions, 
which will also include states that are not proton-and-neutron-even,
are beyond the scope of this paper, 
and  will be ignored.  
To the extent that such excited states are energetically well above
the ground state, this should be a satisfactory approximation.
 

We now see that a nucleon living 
in the self-consistent field of the other nucleons 
inside the \SchiNL~  obeys 
the Dirac equation 
\begin{eqnarray}
0&=&\Big<\Big( i\overrightarrow\partial_\mu \gamma^\mu 
	+\Theta\Big) N\Big>  \\
0&=&\Big< {\overline N}\Big( i\overleftarrow\partial_\mu \gamma^\mu 
	-\Theta\Big) \Big> 
\nonumber
\end{eqnarray}
where 
\begin{eqnarray}
\Theta &\equiv& 
-m^N -\frac{1}{f_\pi^2}{\underline{C^S_{200}}} 
-\frac{1}{f_\pi^2}{\underline{C^V_{200}}} \gamma^0\,, \nonumber 
\end{eqnarray}
with 
\begin{eqnarray}
 {\underline {C^S}} &\equiv& 
 \Big(C^S_{200} -\half {\overline{C^S_{200}}}\Big) \Big<{\overline N}N\Big>
 -\half {\overline{C^S_{200}}} \Big<{\overline N}t_3 N\big>t_3 \\
  {\underline {C^V}} &\equiv& 
 \Big(C^V_{200} -\half {\overline{C^V_{200}}}\Big) \Big<{N^\dagger}N\Big>
 -\half {\overline{C^V_{200}}} \Big<{N^\dagger}t_3 N\Big>t_3 \nonumber \\
 0 &=& \Big[t_3,  {\underline {C^S}}\Big] =  \Big[t_3,  {\underline {C^V}}\gamma^0\Big]  = \Big[t_3, \Theta \Big] \,. \nonumber
 \end{eqnarray}

Ignoring $L_{Static \chi NL}^{4-N;ExcitedNucleon}$,
baryon-number and the third component of isospin are both conserved,
i.e. the associated currents $J^\mu_{Baryon}\equiv{\overline N}\gamma^\mu N$
and 
$J^\mu_{3}\equiv{\overline N}\gamma^\mu  t_3 N$
are both divergence-free.
%
%
%
The neutral axial-vector current 
$J^{5,\mu}_{8}\equiv \frac{\sqrt{3}}{2}{\overline N}\gamma^\mu \gamma^5 N$,
corresponding to the projection onto SU(2)
of the  `eta' NGB $\eta$, part of the unbroken $SU(3)_L\times SU(3)_R$ meson octet,
is also divergence free,
\begin{eqnarray}
\frac{2}{\sqrt{3}}\Big< i\partial_\mu J^{5,\mu}_{8}\Big>
&=&\Big< {\overline N} \Big\{  \Theta,\gamma^5  \Big\} N\Big>
  \\
 &=&2\Big< {\overline N} \Big( -m^N - \frac{1}{f_\pi^2} {\underline {C^S}} \Big) \gamma^5 N\Big>
 \nonumber \\
&\simeq& 0 \,.\nonumber
\end{eqnarray}
This can  be understood as a statement that
the $\eta$ particle 
cannot survive in the parity-even interior of a \SchiNL,
since it is a NGB pseudo-scalar in the chiral limit.  

Similarly, the  axial-vector current 
of the 3rd component of $SU(2)_{L-R}$ isospin 
$J^{5,\mu}_{3} \equiv {\overline N}\gamma^\mu \gamma^5 t_3 N $
is divergence-free,
\begin{eqnarray}
\Big< i\partial_\mu J^{5,\mu}_{3}\Big> 
&=&\Big< {\overline N} \Big\{  \Theta,\gamma^5  \Big\} t_3 N\Big>
 \\
 &=&2\Big< {\overline N} \Big( -m^N - \frac{1}{f_\pi^2} {\underline {C^S}} \Big) \gamma^5 t_3 N\Big>
 \nonumber \\
&\simeq& 0\,, \nonumber
\end{eqnarray}
because the $SU(2)\chi PT$ $\pi_3$ particle 
is also a NGB pseudo-scalar in the chiral limit,
and cannot survive in the interior of a parity-even  \SchiNL.


Even though explicit pion and  $\eta$ fields
vanish in \SchiNL, 
their quantum numbers reappear in its PCAC properties 
from nucleon bi-linears and four-nucleon terms in 
the divergences of axial vector currents.
That these average to zero in \SchiNL~  plays a crucial role in the conservation of axial-vector currents within the liquid.





It is now straightforward to see that, in the liquid approximation, 
a homogeneous $SU(2) \chi PT$ nucleon liquid drop with no meson condensate 
satisfies all relevant CVC and PCAC equations. 
In fact, of all the space-time components of 
the three $SU(2)_{L+R}$ vector currents $J^\mu_a$ 
and three $SU(2)_{L-R}$ axial vector currents $J^{5\mu}_a$, 
only $J^0_3$ does not vanish in \SchiNL.

The neutral $SU(3)_L\times SU(3)_R$ currents are conserved $\big< \partial_\mu J^{\mu}_8 \big>= \big< \partial_\mu J^{5;\mu}_8\big>= 0$
 in the \SchiNL~\\ mean field.  In addition, the neutral $SU(3)_{L+R}$ vector current's spatial components $J^{\mu =1,2,3}_8$ 
and $SU(3)_{L-R}$ axial-vector currents $J^{5;\mu}_8$
all vanish. Only  $J^{0}_8$, proportional to the baryon number density, survives  in the \SchiNL~  mean field.

Since \SchiNL~  chiral nuclear liquids 
satisfy all relevant $\chi PT$ CVC and PCAC equations 
in the liquid phase, 
they are true solutions of the all-orders-renormalized tree level semi-classical liquid equations of motion truncated at ${\mathcal O}(\Lambda_{\chi SB}^0 )$. 

\section{Relation of Static$\chi$NL
	to standard nuclear models}

In a companion paper, 
we apply the Thomas-Fermi approximation to construct
explicit liquid solutions of 
$SU(2)\chi PT$ of protons, neutrons 
and three Nambu-Goldstone boson (NGB) pions. 
Constant-density non-topological solitons, 
i.e. liquids comprised entirely of nucleons, 
emerge as homogeneous and isotropic 
semi-classical static solutions 
at zero  pressure.  
They thus serve as models of the ground state
of both  infinite nuclear matter
and finite liquid drops.
There is no  need for an additional confining interaction
to define the finite-drop surface.
By construction, 
these drops have total spin ${\vec S}= 0$,
 even proton number $Z$, 
 and even neutron number $N$.
We first show symmetric $Z = N$ ground-state 
	zero-pressure Hartree-Fock soliton solutions,
fit to inferred experimental values 
	for symmetric-nuclear-matter 
	density and volume binding energy.
Then, copying nuclides, 
we add $( \frac{Z-N}{Z+N})^2 \ll 1$,
and derive asymmetric $Z \neq N$ nuclear matter,
for which fermion-exchange terms are crucial.
Finally, we show finite 
zero-pressure microscopic liquid drops 
closely resembling the 
Nuclear Liquid Drop Model.
After crude inclusion of electromagnetic chiral symmetry 
breaking, 
our microscopic Static \chiNL~  solitons' saturated nucleon density,
as well as their volume, asymmetry and electromagnetic terms, fit the Bethe-Weizs\"acker Semi-Empirical Mass Formula.
 
This empirical success, 
coupled with the fact that chiral perturbation theory 
is a direct consequence of the Standard Model of particle physics -- 
as correct nuclear physics, atomic physics, {\it etcetera} must ultimately be --
motivates us to consider the connection 
of certain mainstream nuclear-model frameworks 
to the \SchiNL~ solutions we have identified.

\subsection{Density Functionals}

The basic building blocks 
of current relativistic nuclear density functionals \cite{Niksic2008034318} 
are the densities bilinear in the Dirac-spinor field $N$ 
of the nucleon doublet:
\begin{eqnarray}
	\left[\left({\bar N} {\mathcal O}_\tau \gamma_{\mathscr A} N\right)  
	\left({\bar N} {\mathcal O}_\tau\gamma^{\mathscr A}N \right) \right]\,,
\end{eqnarray}
where  
${\mathcal O}_\tau =\left(1, 2{\vec t}\right)$ and 
$\gamma_{\mathscr A} = \left(S,V,T, A, P \right)$.
The nuclear-ground-state density and energy 
 are determined by the self-consistent solution 
 of relativistic linear Kohn-Sham \cite{Kohn19651133} equations. 
 To derive those equations, 
 Niksic et.al. \cite{Niksic2008034318} construct 
 an interaction Lagrangian 
 with four-fermion (contact) interaction terms 
 in the various Lorentz-space isospace channels: 
	scalar-isoscalar $\sigma$ exchange,
	vector-isoscalar $\omega_\mu$ exchange,
	vector-isovector ${\vec \rho}_\mu$ exchange, and
	scalar-isovector ${\vec \delta}$ exchange.
Ignoring explicit electro-magnetic chiral symmetry breaking
\begin{eqnarray}
\label{NiksicLagrangian}
	{\mathcal L}_{Niksic} &=& {\mathcal L}_{Niksic}^{Isoscalar} 
		+ {\mathcal L}_{Niksic}^{Isovector} 
		+ {\mathcal L}_{Niksic}^{Surface}\nonumber \\ 
	{\mathcal L}_{Niksic}^{Isoscalar} 
		&=& {\bar N}\big( i\gamma_\mu \partial^\mu -m_N\big)N \nonumber \\
		&-& \half \alpha_S\Big[\big({\bar N} N\big)  \big({\bar N} N \big) \Big] 
		 -\half \alpha_V\Big[\big({\bar N}\gamma_\mu N\big)  
		 	\big({\bar N} \gamma^\mu N \big) \Big]   \\
	{\mathcal L}_{Niksic}^{Isosvector} 
		&=&-\half \alpha_{TS}\Big[\big({\bar N}2{\vec t} N\big) 
			\cdot \big({\bar N}2{\vec t} N \big) \Big] 
			-\half \alpha_{TV}\Big[\big({\bar N}2{\vec t}\gamma_\mu N\big) 
			\cdot \big({\bar N}2{\vec t} \gamma^\mu N \big) \Big]
	\nonumber \\
	{\mathcal L}_{Niksic}^{Surface} 
		&=&  -\half \delta_S\Big[\partial_\nu 
			\big({\bar N} N\big) \partial^\nu 
			\big({\bar N} N \big) \Big]\,,\nonumber
\end{eqnarray}
where the coefficients are themselves 
functions of the nuclear number density 
normalized to that of nuclear matter:
\begin{eqnarray}
\alpha_S\left(\frac{{N^{\dagger}} N}{{\big[{N^{\dagger}} N\big]}_{Matter}^{Nuclear}}\right),&& 
\alpha_V\left(\frac{{N^{\dagger}} N}{{\big[{N^{\dagger}} N\big]}_{Matter}^{Nuclear}}\right),  \\
\alpha_{TS}\left(\frac{{N^{\dagger}} N}{{\big[{N^{\dagger}} N\big]}_{Matter}^{Nuclear}}\right),&&  
\alpha_{TV}\left(\frac{{N^{\dagger}} N}{{\big[{N^{\dagger}} N\big]}_{Matter}^{Nuclear}}\right)\,.  \nonumber
\end{eqnarray}
In order to be consistent with, and thus legitimately employ,
emergent $Pion-less$\chiPTtw, 
density-functional models must be made to obey
all-orders-renormalized power-counting 
to at least $\Lambda_{\chi{SB}}^{-1}$. 
A beginning would be to re-scale density functional coefficients
to reflect $SU(2)\chi PT$ power counting, and Lorentz invariance, as
\begin{eqnarray}
\alpha_S \Big(\frac{{\overline N}N}{f_\pi^2\Lambda{\chi_{SB}}}\Big),
\alpha_V \Big(\frac{{\overline N}N}{f_\pi^2\Lambda{\chi_{SB}}}\Big),
\alpha_{TS} \Big(\frac{{\overline N}N}{f_\pi^2\Lambda{\chi_{SB}}}\Big),
\alpha_{TV} \Big(\frac{{\overline N}N}{f_\pi^2\Lambda{\chi_{SB}}}\Big)
\end{eqnarray}

Current nuclear density-functional models 
contain non-analytic terms 
inside $\alpha_S, \alpha_V, \alpha_{TS}, \alpha_{TV}$.
These must be made to map onto 
any known non-analytic terms in \chiPTtw~ \cite{Georgi1984}. 

Exchange terms must be included for Hartree-Fock results.

Chhanda Samanta \cite{Samanta2008} claims that ``density functional theory currently predicts long-lived super-heavy elements in a variety of shapes, including spherical, axial and triaxial configurations.  Only when N=184 is approached one expects superheavy nuclei that are spherical in their ground states. Magic islands of extra-stability have been predicted to be around Z=114, 124 or, 126 with N=184, and Z=120, with N=172." If Pionless $SU(2)\chi PT$ confirmed such statements, islands of nuclear stability would move from fantasy to probable fact. 

In the end, 
the requirement that $SU(2)_L\times SU(2)_R \chi PT$ 
have high-accuracy experimental predictive power 
tied to two-massless-quark QCD, 
i.e. the requirement that all chirally invariant terms be included in the Lagrangian and that their coefficients be ``natural",
will force nuclear density-functional theories 
to obey analytic power counting 
to at least ${\mathcal O}\big(\Lambda_{\chi SB}^{-2}\big)$.

\subsection{Nuclear Skyrme models}



A large preexisting class 
of high-accuracy Nuclear Skyrme Models 
	\cite{Nikolaus19921757} 
were first identified by Friar, Madland, and Lynn 
	\cite{Friar19963085, Friar1998145} 
as (almost) derivable from $SU(2)_L\times SU(2)_R ~\chi PT$ liquid
${\mathcal O}\left(\Lambda^n_{\chi SB}\right)$, 
$n = 1, 0, -1, -2$ operators, thus introducing the $\chi PT$ power-counting concept of ``Naturalness" to nuclear Skyrme models. 

Careful and successful comparison of theory to experiment 
for the ground state of 
certain even-even spin-zero spherical closed-shell heavy nuclei 
is a major triumph for 
Relativistic-Mean-Field Point-Coupling Hartree-Fock (RMF-PC-HF) 
"Skyrme" models of nuclear many-body forces 
	\cite{Burvenich2002044308, Finelli2004449, Rusnak1997495, Nikolaus19921757}.
%
For such nuclides, nuclear Skyrme models (almost) obey a much-simplified  \chiPTtw, in which the set of liquid operators is much fewer than the total set of possible non-liquid operators. 

Without prior consideration of chiral liquid $SU(2) \chi PT$, 
Nikolaus, Hoch, and Madland \cite{Nikolaus19921757} 
fit nine coefficients, 
spanning the range of $10^{-4}MeV^{-2}$ to $10^{-18}MeV^{-8}$, 
to the properties of just three heavy nuclei. 
They then predicted the properties 
of another 57 heavy nuclei quite accurately. 
The observational success of their model, 
with the improvements of \cite{Burvenich2002044308} 
is competitive with other nuclear models \cite{deShalit1974}:
binding energies are fit to within $\pm 0.15\%$; 
charge radii are fit to $\pm 0.2\%$;
diffraction radii are fit to $\pm 0.5\%$; 
surface thicknesses are fit to within $\pm 50\%$;
spin-orbit splittings are fit to $\pm 5\%$;
and pairing gaps are fit to $\pm 0.05$MeV. 
The observed isotonic chains, 
fission barriers, etc. are also fit to various high accuracies.

When these 9 coefficients were rescaled \cite{Friar19963085, Friar1998145}, 
as appropriate to $SU(2)_L\times SU(2)_R\chi PT$ liquids, 
these (almost) obeyed $SU(2)\chi PT$ power-counting in $\Lambda_{\chi SB}^{-1}$
through order $\Lambda^{-2}_{\chi SB}$, 
with order-one chiral coefficients. 
A high-accuracy fit 
and its predictions for properties of such heavy nuclei
\cite{Burvenich2002044308} 
showed that
two apparent exceptions have since improved. 
Burvenich, Madland, Marhun, and Reinhard
used 11 coupling constants that, 
when appropriately rescaled with $\Lambda_{\chi SB}$, 
almost obey naturalness \cite{Friar19963085} for $SU(2) \chi PT$ 
	chiral nuclear liquids. 
This obedience of nuclear-Skyrme-model coefficients 
to $\Lambda_{\chi SB}^{-1}$ power-counting \ 
in $SU(2)\chi PT$ 
is now commonly referred to as ``$\chi PT$-naturalness"
in the heavy-nuclear-structure literature \cite{Kortelainen2010}.
Symmetric and asymmetric (finite) nuclear liquid drops 
and bulk nuclear matter in nuclear Skyrme models 
are therefore nearly obedient to $SU(2) \chi PT$.

In order that nuclear Skyrme models emerge with chiral-liquid-like properties within Pionless $SU(2) \chi PT$,
one must carefully consider which operators emerge,
and how their coefficients are related.
But current Skyrme both over-count and omit 
certain liquid operators. 
In order to be consistent with, and thus legitimately employ,
emergent Pionless \chiPTtw, 
they must be made to strictly obey
all-orders-renormalized power-counting 
to ${\mathcal O}(\Lambda_{\chi{SB}}^{-2})$.

\begin{itemize}

\item First re-scale coefficients in (\ref{NiksicLagrangian}) as insisted above
to reflect $SU(2)\chi PT$ power counting, and Lorentz invariance, as
\begin{eqnarray}
\alpha_S \Big(\frac{{\overline N}N}{f_\pi^2\Lambda{\chi_{SB}}}\Big),
\alpha_V \Big(\frac{{\overline N}N}{f_\pi^2\Lambda{\chi_{SB}}}\Big),
\alpha_{TS} \Big(\frac{{\overline N}N}{f_\pi^2\Lambda{\chi_{SB}}}\Big),
\alpha_{TV} \Big(\frac{{\overline N}N}{f_\pi^2\Lambda{\chi_{SB}}}\Big)
\end{eqnarray}

\item 2-Nucleon forces (4-N operators)

\begin{itemize}

\item
For constant $\alpha_V,\alpha_S,\alpha_{TV},\alpha_{TS},\delta_S$, the Lagrangian (\ref{NiksicLagrangian}) is derivable from 2-massless-quark 
QCD, via a Static$\chi NL$ with its Pionless $SU(2)\chi PT$.
				
\item To begin, nuclear Skyrme models should test empirically whether excited nucleon contact terms can really be ignored.
\item If so, replace
\begin{eqnarray}
{\mathcal L}_{Niksic}^{Isosvector} \rightarrow&& \nonumber\\
 {\mathcal L}_{Static\chi NL}^{Isosvector} 
&=& -\half \alpha_{TS}\Big[\big({\bar N}2{t_3} N\big) \cdot \big({\bar N}2{t_3} N \big) \Big] \\
&&-\half \alpha_{TV}\Big[\big({\bar N}2{t_3}\gamma_\mu N\big) \cdot \big({\bar N}2{ t_3} \gamma^\mu N \big) \Big] \nonumber 
\end{eqnarray}
It remains to be seen whether 
	${\mathcal L}_{Static\chi NL}^{Isosvector} $, 
which arises from fermion exchange-terms 
but vaguely resembles 
neutral $\rho^\mu_3$ and $\delta_3$ boson exchange, 
can account, with high accuracy, 
for the known isovector properties 
of nuclear ground-states in nuclear Skyrme models. 
In practice, Niksic {\it et al.} 
neglect the isovector-scalar $\vec \delta$ exchange
(i.e. they set $\alpha_{TS}=0$), arguing that, 
	although the total isovector strength has a relatively well-defined value, 
the distribution between 
	the scalar $\alpha_{TS}$ and vector $\alpha_{TV}$ channels
is not determined by ground-state data.

\item  When ${\mathcal O}\left(\Lambda^0_{\chi SB}\right)$ 
include spinor-interchange and boson-exchange terms,
constant coefficients $\alpha_V,\alpha_S,\alpha_{TV},\alpha_{TS}$
		are linear combinations of 
		$C^S_{200}$, $\overline{C^S_{200}}$, $C^V_{200}$, $\overline{C^V_{200}}$. 
		
\end{itemize}
		
		
	
\item 3-nucleon (6-N operators) and 4-nucleon (8-N operators) contact forces
	
\begin{itemize}
	\item 3-nucleon forces, of order  
			${\mathcal O}\left(\Lambda^{-1}_{\chi SB}\right)$ + ${\mathcal O}\left(\Lambda^{-2}_{\chi SB}\right)$,
			are smaller than 
			${\mathcal O}\left(\Lambda^{0}_{\chi SB}\right)$
			2-nucleon forces \cite{Weinberg1990288, Weinberg19913, Weinberg1992114}. 
			\item 4-nucleon 
			${\mathcal O}\left(\Lambda^{-2}_{\chi SB}\right)$ forces 
			are smaller still than 3-nucleon 
			${\mathcal O}\left(\Lambda^{-1}_{\chi SB}\right)$ forces \cite{Weinberg1990288}, \cite{Weinberg19913}, \cite{Weinberg1992114}. 
			Including  separate $C^S_{400}$ and $C^V_{400}$ 
			over-counts independent chiral coefficients. 

			\item For example, to ${\mathcal O}(\Lambda
_{\chi_{SB}} ^{-2})$,
\begin{eqnarray}
\alpha_V \Big(\frac{{\overline N}N}{f_\pi^2\Lambda{\chi_{SB}}}\Big)&\simeq& \alpha_V(0) 
+\frac{1}{3} C_{300} \Big[\frac{{N^\dagger}N}{f_\pi^2\Lambda{\chi_{SB}}}\Big] 
+\frac{1}{4} C_{400} \Big[\frac{{N^\dagger}N}{f_\pi^2\Lambda{\chi_{SB}}}\Big]^2 \nonumber \\
\alpha_S \Big(\frac{{\overline N}N}{f_\pi^2\Lambda{\chi_{SB}}}\Big)&\simeq& \alpha_S(0) 
\end{eqnarray}
representing 2-nucleon, 3-nucleon and 4-nucleon contact terms respectively. Since non-relativistic  $N^{\dagger}N$ and ${\overline N}N$ differ by relativistic corrections of ${\mathcal O}(\Lambda
_{\chi_{SB}} ^{-2})$, $SU(2)\chi PT$ requires $\alpha_V(0), C_{300}, C_{400}$ to be $\simeq {\mathcal O}(1)$ natural.

\item Spinor-interchange terms must also be added to consistently preserve Hartree-Fock quantum loop power counting.

\end{itemize}

	\item Incorporate
		${\mathcal O}\left(\Lambda^{-2}_{\chi SB}\right)$  
		nuclear-surface terms.
		Because they only involve differentials 
		of the baryon number density, 
		only certain surface terms are invariant 
		under local $SU(2) \chi PT$ transformations, 
		and do not contribute to $SU(2)_{L+R}$ or $SU(2)_{L-R}$ currents 
		affecting CVC or PCAC properties. 
		These terms 
		replace the scalar $\sigma$ particle in the Chin-Walecka model 
		in describing the nuclear surface \cite{Chin197424, Serr197810, Serot19861}.  
		The surface term must be re-scaled
\begin{eqnarray}
{\mathcal L}_{Niksic}^{Surface} &\rightarrow& {\mathcal L}_{Static \chi NL}^{Surface} = -\half C_{220}\Big[ \frac{\partial_\nu}{\Lambda{\chi_{SB}}} \big({\bar N} N\big) \frac{\partial^\nu}{\Lambda{\chi_{SB}}} \big({\bar N} N \big) \Big]\,.\,
\end{eqnarray}
with constant $C_{220} \simeq {\mathcal O}(1)$ in order to obey naturalness and absorb all-orders quantum loops. ${\mathcal L}_{Static \chi NL}^{Surface}$ is invariant under $SU(2)_L\times SU(2)_R$ transformations, including pions, but is automatically pion-less, even without the liquid approximation. It contains no dangerous $\partial_0 \sim m_N$ nucleon mass terms, so non-relativistic re-ordering is un-necessary. 
Nucleon-exchange and spinor-interchange interactions must also be included.

		\item According to strict ${\mathcal O}\left(\Lambda^{-2}_{\chi SB}\right)$ 
 power counting \cite{Weinberg1990288, Weinberg19913, Weinberg1992114}, current nuclear Skyrme models 
 sometimes over-count chiral-liquid operators.
 For example, to ${\mathcal O}\left(\Lambda^{-2}_{\chi SB}\right)$, 
  only eight of eleven coupling constants 
 	considered in \cite{Burvenich2002044308} 
	are truly independent.
	\item   The authors of \cite{Burvenich2002044308} are also missing chiral-liquid operators
where non-relativistic re-ordering of the large nucleon mass term $\sim \partial_0$ is necessary.
Such time-dependent operators may be important for high-accuracy nuclear structure, 
and will also affect ordinary heavy nuclear 
$SU(2)_{L+R}$ and $SU(2)_{L-R}$ currents, 
CVC and PCAC. 

\end{itemize} 

\subsubsection{{\it Ab Initio} calculations}

 It is beyond the scope of this paper 
 to construct a complete minimal 
 ${\mathcal O}\left(\Lambda^{-2}_{\chi SB}\right)$
 set of chiral-liquid operators 
 for nuclear Skyrme models, 
 but a systematic program of calculation 
 of detailed properties of the ground state 
 of even-even spin-zero spherical closed-shell heavy nuclei 
 in RMF-PC-HF (and nuclear liquid drops) 
 with that set 
 is necessary in order to extract predictions for nuclear structure 
 from \SchiNL~  emergent from $SU(2) \chi PT$. 
 
Going forward, it is important to understand whether the contribution of
\begin{eqnarray}
\label{ExcitedNucleonAppendix}
&& - L_{Static \chi NL;\quad t_\pm t_\mp}^{4-N;ExcitedNucleon} = \half \sum_{\Psi \neq Static\chi NL} \frac{1}{ f_\pi^2}\\
 &&\quad  \times \Big[ C^{T=1}_{S} \brao ( \overline{N_c^\alpha}(t_\pm)_{cd} N_d^\alpha)\big\vert \Psi \big> \big< \Psi \big\vert
( \overline{N_e^\lambda}(t_\mp)_{ef}  N_f^\lambda)\keto \nonumber \\
 &&\quad \,\, + C^{T=1}_{V} \brao ( \overline{N_c^\alpha}(t_\pm)_{cd}\gamma^{\mu;\alpha\beta} N_d^\beta)\big\vert \Psi \big> \big< \Psi \big\vert
( \overline{N_e^\lambda}(t_\mp)_{ef}\gamma_{\mu}^{\lambda\sigma} N_f^\sigma)\keto \Big]\nonumber
\end{eqnarray}
is numerically material to empirical models. Such terms involve proton-odd neutron-odd intermediate states, and may require explicit pion-exchange effects lying outside pion-less $SU(2)\chi PT$, thus significantly complicating {\it Ab Initio} calculations.


If Nuclear Skyrme Models properly incorporating strict $SU(2)_L\times SU(2)_R$ power-counting to 
${\mathcal O}\left(\Lambda^{-2}_{\chi SB}\right)$ in chiral liquids, and possibly including higher representations such as the $\Delta (1232)$, 
were to capture empirical reality, including no-core shell structure,  to high accuracy for those nuclear-ground-states which are to be regarded as liquid drops, that success would have been traced directly to the global symmetries of 2-massless-quark QCD.

\subsection{Neutron Stars}
Putting aside exotica (i.e. quark condensates, strange-kaon condensates, etc.), 
we conjecture that much of the structure of neutron stars may be traced directly to 2-massless-quark QCD, and thus directly to the Standard Model. 

The models of Harrison \& Wheeler \cite{Harrison1958}, Salpeter \cite{Salpeter1961669} and  Baym, Pethic and Sutherland \cite{Baym1971299}, 
all based on the Bethe-Weizs\"acher Semi-Empirical Mass Formula
\cite{Shapiro1983}, 
would seem to be implied by our companion paper \cite{Lynn2018}. 
If Density Functional and Skyrme models 
can be modified to strictly obey $SU(2)\chi PT$, highly credible and predictive Standard-Model neutron-star structure would follow. 

Kim et. al. \cite{Kim2018} have recently used 
gravitational wave observations 
\cite{Abbott2017161101},
density functional technology, and reasonable constraints on Skyrme models 
(from stable nuclei, nuclear matter and the maximum mass of neutron stars), 
to constrain the tidal deformities in single neutron stars and binaries. 
If such constraints could be traced directly to QCD and the SM as conjectured here,
a strong new connection between General Relativity and the Standard Model
will have appeared.

\subsection{2-light-mass-quark QCD's $SU(2)\chi PT$ symmetry-breaking terms}

B.W. Lynn\cite{Lynn1993281} first introduced the idea that $SU(2)\chi PT$ could admit a liquid phase.
His Lagrangian included only $SU(2)\chi PT$ terms of ${\mathcal O}(\Lambda_{\chi SB})$ and ${\mathcal O}(\Lambda_{\chi SB}^0)$ and ignored electro-magnetic breaking.
These included strong-interaction terms which survive the chiral limit, as well as 
explicit chiral symmetry breaking terms which do not. 
But he was careful to include only/all those terms which survive the approximate Static$\chi NL$ dynamical symmetries discussed in this paper.

The symmetry-breaking terms have $m=0, l=1, n=1$ in
(\ref{LchiPTfull}). Ignoring $\pi^\pm - \pi^0$ mass splitting, these are
\begin{eqnarray}
\label{SymmetryBreaking}
L_{\chi PT}^{N;\chi SB} &\simeq& \big[ m_{up}+m_{down}\big]\big[a_1+a_2+a_3\big]\Big[1-\cos{\frac{2\pi}{f_\pi}}]  \nonumber \\
(a_1,a_2,a_3;m_{up},m_{down}) &=& (0.28, -0.56, 1.3\pm 0.2;6 MeV,12 MeV)\,, 
\end{eqnarray} 
with constants measured in $SU(3)_L\times SU(3)_R \chi PT$ processes 
\cite{Georgi1984} and \cite{Lynn2010}.

Since $L_{\chi PT}^{N;\chi SB} > 0$, the symmetry-breaking terms have the effect of lowering the effective nucleon mass inside a static $\pi = \vert {\vec\pi} \vert$ condensate. \cite{Lynn1993281}
showed that an unphysical `pion-nucleon coupling' $\beta \sigma_{\pi N} \geq 400 MeV$, with $\beta \geq 6.66$, causes an S-wave ${\vec \pi}^2$ pion condensate to form. He also showed that
the experimental values $\beta=1;\sigma_{\pi N} \simeq 60 MeV$ allow no such S-wave condensate to form in ordinary heavy nuclei, in agreement with observation \cite{Ericson1988}, and thus avoided the disaster of ``parity doubling\cite{Lynn1993281}" in the chart of the nuclides.

{Note that  (\ref{SymmetryBreaking}) is further suppressed in the 2-light-quark sector because 
\begin{eqnarray}
(m_{up}+m_{down}) \sim \frac{m_\pi^2}{\Lambda{\chi SB}}\sim 0.02 GeV
\end{eqnarray}
which is why un-physically large  $\beta$ was necessary to form the un-physical $\pi$-condensate.
Maybe this also explains why it is empirically successful to take certain nuclear structure to be independent of the 
pion mass \cite{vanKolckSMat50}.

We conjecture that $Pion-less$ $SU(2)\chi PT$ of Static$\chi NL$ for certain nuclides can be shown to effectively include all ${\mathcal O}(\Lambda_{\chi SB})$ and ${\mathcal O}(\Lambda_{\chi SB}^0)$ non-strange power-counting terms, both those from the chiral-limit and those from $m_{up},m_{down}\neq 0$ chiral symmetry breaking.}
\footnote{
{Strange Chiral Nuclear Liquids \cite{Lynn2010}, a form of Strange Baryon Matter \cite{Lynn1990186}, consist of a Static${\chi NL}$ immersed in a kaon condensate driven by large $m_{strange}\simeq 0.24 GeV,\beta\simeq 9$. These strange chiral liquids are identified \cite{Lynn2010}, \cite{Coffey2018a} as a possible $SU(3)_L\times SU(3)_R \chi PT$ MACRO dark matter candidate \cite{Lynn2018} non-topological soliton \cite{Coffey2018b} which is fully consistent with the dynamics of ordinary nuclides in this paper. 
}
}
\section{Conclusions}
{The Standard Model of particle physics, augmented by neutrino mixing and General Relativity (i.e. Frank Wilczek's ``Core Theory" \cite{Wilczek2016})
is the most powerful, accurate, predictive, successful and experimentally successful scientific theory known to humans.
No experimental counter-example has ever been observed in the known universe.
Its local $SU(3)_{Color}$ Quantum Chromo-Dynamic subset is,
according to all experimental evidence, 
the complete and correct theory of the  strong interactions
of known fundamental particles 
at all energies accessible to current technology
\footnote{
   	Neutrinos
	may have undiscovered interactions connected to their
	mass and to their flavor oscillations.
   	These are unlikely to affect the conclusions of this paper.		
}
It must therefore underlie the complete and correct theory of the structure
and interactions of atomic nuclei.}

In this paper, we have explored 
some of the implications of this inescapable connection
for nuclear structure as directly derivable from Standard Model, 
especially from the global symmetries of QCD.
In this we have been 
guided by two key observations: 
that nuclei are made of protons and neutrons,
not quarks;
and that the  up and down quarks, 
which are  the fermionic constituents of the protons and neutrons,
are much lighter than 
the principal mass scales of QCD, 
such as the proton and neutron masses.
Taken together, 
these strongly suggest that the full complexity of the Standard Model
can largely be captured, for the purposes of nuclear physics, 
by an effective field theory (EFT) --
$SU(2)_L \times SU(2)_R$ chiral perturbation theory (\chiPTtw)
of protons and neutrons.

In writing down an EFT Lagrangian, 
one incorporates all analytic higher-order quantum-loop corrections 
into tree-level amplitudes.
\chiPTtw~ enables the operators of that EFT Lagrangian 
(and the states) 
to be expressed
as a perturbation expansion in inverse powers of the
chiral-symmetry-breaking scale 
$\Lambda_{\chi SB}\simeq 1\GeV$.

Building on this longstanding insight, 
we have studied the chiral limit of \chiPTtw~ EFT, 
including only operators of order 
$\Lambda_{\chi SB}$ and $\Lambda_{\chi SB}^{0}$.
We find that
\chiPTtw~ of  protons,\footnote{
	Note that in the chiral limit, electromagnetic interactions are ignored.
	} 
neutrons and 3 Nambu-Goldstone boson (NGB) pions - 
admits a semi-classical liquid phase,
a Static Chiral Nucleon Liquid (\SchiNL).

\SchiNL s~ are made entirely of nucleons,
with zero anti-proton and anti-neutron content.
They are parity even and time-independent.
As we have studied them so far, 
not just the total nuclear spin ${\vec S}= 0$, 
but also the local expectation value for spin $< {\vec s}> \simeq 0$.
Similarly, 
the nucleon momenta vanish locally in the \SchiNL~ rest frame.
For these reasons, 
our study of \SchiNL~ is applicable 
to bulk ground-state spin-zero nuclear matter,
and to the ground state of appropriate spin-zero parity-even nuclei 
with an even number $Z$ of protons and an even number $N$ of neutrons.


We classify these solutions of \chiPTtw~ as ``liquid" because
energy is required both to pull the constituent nucleons further apart
and to push them closer together.
This is analogous with the balancing of the
attractive Lorentz-scalar $\sigma$-exchange force 
and 
the repulsive Lorentz-vector $\omega_\mu$-exchange force 
in the Walecka model.
The nucleon number density therefore takes a saturated value
even in zero external pressure (e.g. in the absence of gravity),
so are not a ``gas."
Meanwhile they are statistically homogeneous and isotropic,
lacking the reduced symmetries of crystals or other solids.


We have shown that in this ground-state liquid phase, 
the expectation values 
of many of the allowed operators 
of the most general \chiPTtw~ EFT Lagrangian
vanish or are small.
We have further conjectured that, 
for studying (static) ground-state systems,
many more operators are small 
because they involve transitions to excited intermediate states. 
Going forward, 
it is imperative to understand the contribution 
of $L_{Static \chi NL}^{4-N;ExcitedNucleon}$
\eqref{ExcitedNucleon}
to empirical models of nuclear ground states.

We have also shown that this ground-state liquid phase 
does not support a classical pion field --
infrared pions decouple from these solution.
We expect that this emergence of ``pion-less \chiPTtw''
is at the heart of the apparent theoretical independence 
of much successful nuclear physics from pion properties 
such as the pion mass.

In a companion paper, 
we will use the Thomas-Fermi approximation to provide 
an explicit  ``proof of principal" solution 
for a liquid phase of \chiPTtw,
{\it i.e.}~\SchiNL~  non-topological solitons. 
We demonstrate there the existence 
of zero-pressure non-topological soliton \SchiNL s, 
with both macroscopic (infinite nuclear matter) 
and microscopic (nucleide ground states).

We conjecture that, for appropriate nuclides, proper inclusion of $1/\Lambda_{\chi SB}$ 
and $\left(1/\Lambda_{\chi SB}\right)^2$ \\ 
  \chiPTtw~ operators 
will result in accurate "natural" nuclear Skyrme models, 
exhibiting ``no-core" shell structure, 
with approximate \SchiNL~  structure.

We speculate that the extension of the line of thinking
contained in this paper to  $SU(3)_L\times SU(3)_R$\chiPT, 
will be instructive on the experimentally current question 
of strange nuclei, 
and on the astrophysically relevant question of strange nuclear matter.

{The Standard Model (augmented by neutrino mixing) is, 
as a  result of five decades of experimental and theoretical effort,
a remarkably complete and correct description 
of all non-gravitational interactions of known fundamental particles, 
without experimentally identified exception. 
Nature has been kind, by building atoms out of electrons and nuclei,
and nuclei out of protons and neutrons, 
and by making the up and down quark so much lighter than those,
to afford us a possible pathway to relate the emergent physics of:
atoms \footnote{Gary Feinberg: private communication}; the deuteron \cite{Weinberg1965672, vanKolck19942932}; the heavy nuclides in this paper; and the structure of the proton \cite{NChristSMat50} in lattice gauge theory, directly to the fundamental interactions of the Standard Model.
It is incumbent on us to avail ourselves of that kindness
by striving to obediently connect our phenomenological/empirical models to
Nature's magnificent fundamental theory.}

\section*{Acknowledgments}

This paper is dedicated to BWL's teacher, 
Gerald Feinberg (1933-1992) who predicted the muon neutrino; and 
teased emergent physics - positronium, 
muonium (i.e. renormalizable bound states), the periodic table (i.e. atomic physics), and the polarizability of di-hydrogen (i.e. chemistry) - directly from the Standard Model.

We thank Brian J. Coffey (BJC) for technical assistance.
GDS is partially supported by a grant from the US DOE 
to the particle-astrophysics theory group at CWRU. 
We thank Kellen McGee and David Jacobs for their efforts
directed at the follow-up paper.
BWL and BJC thank Chien-Shiung Wu for teaching us nuclear physics, and Robert Serber for many enlightening discussions of his work on nuclear structure.

\bibliographystyle{h-physrev}
\bibliography{ChiralNuclearLiquids}
\begin{appendices}
\numberwithin{equation}{section}

\section{$SU(2)_L\stimes{}SU(2)_R$ $\chi PT$ of a nucleon doublet
and a {pion}
triplet in the chiral limit}
\label{app:chiptLagrangian}

The chiral symmetry of two light quark flavors in QCD, 
together with the symmetry-breaking and Goldstone's theorem, 
makes it possible to obtain an approximate solution to QCD at low energies 
using a $SU(2)_L \times SU(2)_R$ EFT,
where the degrees of freedom are hadrons \cite{
	Weinberg19681568, Weinberg1978327, Coleman19692239, Callan19692247, 
	Gasser1984142, Gasser1985465, Manohar1984189, Georgi1993187, Georgi1984}.   
In particular, the non-linear \chiPTtw~ effective Lagrangian 
has been shown to successfully model the interactions of pions with nucleons,
where a perturbation expansion 
(e.g., in soft momentum $\vec{k}/\Lambda_{\chi SB} \ll 1$,  
baryon number density $\frac{N^\dagger N}{f^2_\pi \Lambda_{\chi SB}} \ll 1$, 
for chiral symmetry breaking scale $\Lambda_{\chi SB} \approx 1$ GeV) 
has demonstrated predictive power. 
Power-counting in $\Lambda_{\chi SB}^{-1}$ 
includes all analytic quantum-loop effects 
into experimentally measurable coefficients of $SU(2)_L \times SU(2)_R$ 
current-algebraic operators 
obedient to the global symmetries of QCD, 
with light-quark masses generating additional explicit 
chiral-symmetry-breaking terms. 
Therefore, \chiPTtw~ tree-level calculations 
with a power-counting effective Lagrangian 
are to be regarded as true predictions of QCD 
and the Standard $SU(3)_C\times SU(2)_L\times U(1)_Y$ Model 
of elementary particles.

\subsection{Non-linear transformation properties}

We present the Lagrangian of unbroken \chiPTtw~ 
of a nucleon doublet and a pseudo-Nambu-Goldstone-Boson (pNGB) triplet. 
We employ the defining SU(2) strong-isospin representation of
unitary $2\stimes2$ Pauli matrices $\sigma_a$, 
with asymmetric structure constants 
$f_{abc}=\epsilon_{abc}$
\begin{equation}
  \begin{aligned}
 	\label{SUthreealgebra}
    t_a = \frac{\sigma_a}{2}\,, \quad a = 1,3 \\
    \Tr(t_a t_b) = \frac{\delta_{ab}}{2} \\
    \left [ t_a, t_b \right ] = if_{abc}t_c \\
    \left \{t_a, t_b \right \} = \frac{\delta_{ab}}{2}     \,.
  \end{aligned}
\end{equation} 
The  $SU(2)_{L+R}$ vector and $SU(2)_{L-R}$ axial-vector charges 
obey the algebra
\begin{equation}
  \begin{aligned}
  \label{VandAchargesalgebra}
    \left [ Q^{L+R}_{a}, Q^{L+R}_{b} \right ] = if_{abc}Q^{L+R}_c \\
    \left [ Q^{L-R}_{a}, Q^{L-R}_{b} \right ] = if_{abc}Q^{L+R}_c \\
    \left [ Q^{L+R}_{a}, Q^{L-R}_{b} \right ] = if_{abc}Q^{L-R}_c \,.
  \end{aligned}
\end{equation} 

We consider a triplet representation of NGBs, 
\begin{equation}
  \begin{aligned}
 \label{pGoctetrepn1}
    \pi_at_a = \frac{1}{\sqrt{2}}\left [
    \begin{tabular}{ccc} 
    $\frac{\pi^0}{\sqrt{2}}$ & $\pi^+$ \\ 
    $\pi^-$ & $-\frac{\pi^0}{\sqrt{2}}$      \\
    \end{tabular} \right ]
  \end{aligned}
\end{equation}
and a doublet of nucleons,
\begin{equation}
  \begin{aligned}
  \label{Boctetrepn}
    N = \left [
    \begin{tabular}{ccc} 
     $p$ \\ 
     $n$ \\
    \end{tabular} \right ]\,.
  \end{aligned}
\end{equation}

For pedagogical simplicity, 
representations of higher mass are neglected, 
even though the \\ $SU(3)_L\times SU(3)_R$ baryon decuplet 
(especially $\Delta_{1232}$) 
is known to have important nuclear structure \cite{Ericson1988} 
and scattering \cite{Jenkins2002242001} effects.

Since \chiPTtw~ matrix elements 
are independent of representation \cite{Coleman19692239, Callan19692247},
we choose a representation \cite{Manohar1984189, Georgi1993187, Georgi1984} 
where
  the NGB octet has only derivative couplings, 
\begin{equation}
	\label{Sigmarepn}
  \Sigma \equiv \expon(2i \pi_a \frac{t_a}{f_\pi})\,.
\end{equation}
Under a unitary global $SU(2)_L\stimes{}SU(2)_R$ transformation, 
given by $L\equiv \exp(il_at_a)$ and $R \equiv \exp(ir_at_a)$,
\begin{equation}
\label{Sigmatransfn}
\Sigma \rightarrow \Sigma' = L\Sigma R^\dagger\,.
\end{equation}
It also proves useful to introduce the  ``square root'' of $\Sigma$
\begin{equation}
  \xi \equiv  \expon(i \pi_a \frac{t_a}{f_\pi})\,,
\end{equation}
which transforms as
\begin{eqnarray}
    \xi &\rightarrow& \xi^\prime =  \expon(i\pi^\prime_a\frac{t_a}{f_\pi}) \nonumber\,.
\end{eqnarray}
We observe that
\begin{eqnarray}
    \xi^\prime &=& L\xi U^\dagger = U\xi R^\dagger \,,
\end{eqnarray} 
for some unitary local transformation matrix $U(L,R,\pi_a(t,x))$.

The vector and axial-vector NGB currents
\begin{equation}
  \begin{aligned}
    V_\mu \equiv \frac{1}{2}(\xi^\dagger\partial_\mu\xi + \xi\partial_\mu\xi^\dagger) \\
    A_\mu \equiv \frac{i}{2}(\xi^\dagger \partial_\mu \xi - \xi\partial_\mu\xi^\dagger)
  \end{aligned}
\end{equation}
transform straightforwardly as
\begin{equation}
  \begin{aligned}
    V_\mu \rightarrow V^\prime = UV_\mu U^\dagger + U \partial_\mu U^\dagger \\
    A_\mu \rightarrow A^\prime = U A_\mu U^\dagger \,.
  \end{aligned}
\end{equation}
Meanwhile the nucleons transform as
\begin{equation}
    N \rightarrow N^\prime = UN 
\end{equation}
and
\begin{equation}
    D_\mu N \equiv \partial_\mu N + V_\mu N
        \rightarrow U(D_\mu N) \,.
\end{equation}

\subsection{$\Lambda_{\chi SB}$ power counting}

The $SU(2)\chi PT$ Lagrangian,
including all analytic quantum-loop effects 
for soft momenta ($\ll1$GeV) \cite{Manohar1984189, Georgi1993187}, 
can now be written:
\begin{eqnarray}
\label{LchiPTfull}
  L_{\chi PT} &=&  \\
     &&\!\!\!\!\!\!\!\!\!\!\!\!\!\!\!\!\!\!\!\!\!\!\!\!\!
     -\!\!\!\!\!\sum_{\substack{l,m,n\\l+m\geq1}}\!\!\!
     	 C_{lmn} f^2_\pi\Lambda^2_{\chi SB}
     \left ( \frac{\partial_\mu}{\Lambda_{\chi SB}}\right )^{\!\!m}
     \left ( \frac{\overline{N}}{f_\pi\sqrt{\Lambda_{\chi SB}}}\right )^{\!\!l} 
     \left ( \frac{N}{f_\pi\sqrt{\Lambda_{\chi SB}}}\right )^{\!\!l} 
     \left ( \frac{m_{quark}}{\Lambda_{\chi SB}}\right )^{\!\!n} 
     f_{lmn}\left ( \frac{\pi_a}{f_\pi} \right ) \nonumber\,,
\end{eqnarray}
where $f_{lmn}$ is an analytic function,
and the dimensionless constants $C_{lmn}$ 
are ${\mathcal O}(\Lambda_{\chi SB}^0)$ and, presumably,
$\sim 1$. 
As a power series in $\Lambda_{\chi SB}$,
\begin{equation}
\label{LchiPTpowers}
  L_{\chi PT} \sim \Lambda_{\chi SB} + (\Lambda_{\chi SB})^0 + \frac{1}{\Lambda_{\chi SB}} + \left ( \frac{1}{\Lambda_{\chi SB}} \right )^2 +... 
\end{equation}

We take, self-consistently, $\Lambda_{\chi SB}\simeq 1 GeV$
and, in higher orders,
reorder the non-relativistic perturbation expansion in $\partial_0$ 
to converge with large nucleon mass 
$m^N \approx \Lambda_{\chi SB}$ \cite{Weinberg1990288, Weinberg19913, Weinberg1992114}.
As the terms in \eqref{LchiPTfull} already include all loop corrections,
we can perform tree-level calculations 
to arrive at strong-interaction predictions.

\subsection{The Chiral Limit}

For the purposes of this paper, 
we retain from \eqref{LchiPTfull} only terms of order 
$\Lambda_{\chi SB}$ and $\Lambda^0_{\chi SB}$, i.e. $1\leq m+l+n\leq2$.
We can further divide $L_{\chi PT}$
into a symmetric piece 
  (i.e., spontaneous $SU(2)_{L-R}$ breaking with massless Goldstones) 
  and a symmetry-breaking piece 
  (i.e., explicit $SU(2)_{L-R}$ breaking, traceable to quark masses) 
  generating three massive pNGB: 
\begin{equation}
	L_{\chi PT} = L^{Symmetric}_{\chi PT} + L^{Symmetry-Breaking}_{\chi PT} \,.
\end{equation}
In this paper, 
we are interested only in unbroken \chiPTtw~ 
and so take $n=0$ in \eqref{LchiPTfull}
\begin{eqnarray}
	L^{Symmetry-Breaking}_{\chi PT} = 0.
\end{eqnarray}

We separate $L^{Symmetric}_{\chi PT}$ into 
pure-meson terms, 
terms quadratic in baryons (i.e. nucleons), 
and four-baryon terms:
\begin{eqnarray}
	L^{Symmetric}_{\chi PT} &=& 
	L^{\pi;Symmetric}_{\chi PT} + L^{N;Symmetric}_{\chi PT} 
	+ L^{4-N;Symmetric}_{\chi PT} 
\end{eqnarray}
with (as in \eqref{SymmetricLagrangian})
\begin{eqnarray}
\label{SymmetricLagrangianAppendix}
	L^{\pi; Symmetric}_{\chi PT} &=& 
		\frac{f^2_\pi}{4}\Tr \partial_\mu \Sigma \partial^\mu \Sigma^\dagger 
		+ L^{\pi;Symmetric}_{\chi PT;non-Analytic}\\
	L^{N; Symmetric}_{\chi PT} &=& 
		\overline{N} \left(i\gamma^\mu D_\mu -  \tildemone\right)N 
		- g_A\overline{N}\gamma^\mu\gamma^5{A_\mu N }  \nonumber\\
	L^{4-N;Symmetric}_{\chi PT} &\sim& 
		\frac{1}{f^2_\pi}\left(\overline{N}\gamma^{\mathscr A} N\right)
			\left( \overline{N}\gamma_{\mathscr A} N\right) +++\nonumber\,,
\end{eqnarray}

As described below \eqref{SymmetricLagrangian},
the parentheses in the four-nucleon Lagrangian 
indicate the order of $SU(2)$ index contraction,
and  $+++$  indicates that one should include 
all possible combinations of such contractions.
As usual,
  $\gamma^{\mathscr A} \equiv
  \left(1, \gamma^\mu,  i\sigma^{\mu\nu},  
  i\gamma^\mu \gamma^5,  \gamma^5\right)$,
for ${\mathscr A}=1,...,16$
(with 
$\sigma^{\mu\nu} \equiv \frac{1}{2}[\gamma^\mu,  \gamma^\nu ]$).
These are commonly referred to as 
scalar (S), vector (V), tensor (T), axial-vector (A), and pseudo-scalar (P)
respectively. 

\subsection{$SU(2)_L\times SU(2)_R$ invariant 4-nucleon contact interactions}
Focus on the 4-fermion terms in \eqref{SymmetricLagrangianAppendix}.

Using the completeness relation for $2\times 2$ matrices (sum over $A=0,3$)
\begin{eqnarray}
\sigma^B &=& (1, {\vec \sigma});
\quad  \delta_{cf}\delta_{ed} = 
	\frac{1}{4}\sum_{B=0}^3\sigma^B_{cd} \sigma^B_{ef}\,.
\end{eqnarray}
(We use $\alpha ...\sigma$ for relativistic spinor indices, 
while $a...f$ are isospin indices.)
Both iso-scalar and iso-vector 4-nucleon contact interactions 
appear in the $SU(2)_L\times SU(2)_R$ invariant Lagrangian:
\begin{eqnarray}
	\label{L4NSymmetricchiPT}
	L^{4-N;Symmetric}_{\chi PT} &=& \frac{1}{f^2_\pi}C^{T=0}_{\mathscr A}( \overline{N_a^\alpha}\gamma^{{\mathscr A}\alpha\beta} N_a^\beta)( \overline{N_b^\lambda}\gamma_{\mathscr A}^{\lambda\sigma} N_b^\sigma) \nonumber\\
	&&\quad\quad + \frac{1}{f^2_\pi}C^{T=1}_{\mathscr A}( \overline{N_a^\alpha}\gamma^{{\mathscr A}\alpha\beta} N_b^\beta)( \overline{N_b^\lambda}\gamma_{\mathscr A}^{\lambda\sigma} N_a^\sigma)  \nonumber\\
	&&\!\!\!\!\!\!\!\!\!\!\!\!\!\!\!\!\!\!\!\!\longrightarrow 
		\frac{1}{f^2_\pi}C^{T=0}_{\mathscr A}( \overline{N_c^\alpha}U^\dagger_{ca}\gamma^{{\mathscr A}\alpha\beta} U_{ad}N_d^\beta)( \overline{N_e^\lambda}U^\dagger_{eb}\gamma_{\mathscr A}^{\lambda\sigma} U_{bf}N_f^\sigma) \\
	&&\quad\quad + 
		\frac{1}{f^2_\pi}C^{T=1}_{\mathscr A}( \overline{N_c^\alpha}U^\dagger_{ca}\gamma^{{\mathscr A}\alpha\beta} U_{bd}N_d^\beta)( \overline{N_e^\lambda}U^\dagger_{eb}\gamma_{\mathscr A}^{\lambda\sigma} U_{af}N_f^\sigma) \nonumber \\
	&&\!\!\!\!\!\!\!\!\!\!\!\!\!\!\!\!\!\!\!\!= 
		\frac{1}{f^2_\pi}C^{T=0}_{\mathscr A}( \overline{N_c^\alpha}\gamma^{{\mathscr A}\alpha\beta} N_c^\beta)( \overline{N_e^\lambda}\gamma_{\mathscr A}^{\lambda\sigma} N_e^\sigma) \nonumber\\
		&&\quad\quad + 
		\frac{1}{f^2_\pi}C^{T=1}_{\mathscr A}( \overline{N_c^\alpha}\gamma^{{\mathscr A}\alpha\beta} N_d^\beta)( \overline{N_e^\lambda}\gamma_{\mathscr A}^{\lambda\sigma} N_f^\sigma)
		\delta_{cf}\delta_{ed} \nonumber \\
	&&\!\!\!\!\!\!\!\!\!\!\!\!\!\!\!\!\!\!\!\!= 
		\frac{1}{f^2_\pi}C^{T=0}_{\mathscr A}( \overline{N_c^\alpha}\gamma^{{\mathscr A}\alpha\beta} N_c^\beta)( \overline{N_e^\lambda}\gamma_{\mathscr A}^{\lambda\sigma} N_e^\sigma) \nonumber \\
        &&\quad\quad+
		\frac{1}{4f^2_\pi}  \sum _{B=0}^3  C^{T=1}_{\mathscr A}( \overline{N_c^\alpha}\sigma^B_{cd}\gamma^{{\mathscr A}\alpha\beta} N_d^\beta)( \overline{N_e^\lambda}\sigma^B_{ef}\gamma_{\mathscr A}^{\lambda\sigma} N_f^\sigma)\,. \nonumber
\end{eqnarray}

\subsection{Non-analytic NGB pion interactions}
Non-analytic interactions of pions 
are induced in quantum loops. 
There are situations where loop effects are important 
and can be qualitatively distinguished from tree-level interactions 
by their analytic structure. 
For example, 
the $\pi_a + \pi_b \rightarrow \pi_c + \pi_d$ scattering amplitude 
contains a term \cite{Georgi1984} in the chiral limit.
\begin{eqnarray}
\label{LpiSymmetricnonAnalytic}
	L^{\pi; Symmetric}_{non-Analytic}&\leftrightarrow& 
		\Big[ -\delta_{ab}\delta_{cd}\frac{s^2}{32\pi^2} 
			 - \delta_{ac}\delta_{bd}\frac{3s^2+u^2-t^2}{196\pi^2} 
	 \\&-& 
	\delta_{ad}\delta_{bc}\frac{3s^2+t^2-u^2}{196\pi^2} \Big] 
	\ln\big(-\frac{s}{\kappa} \big)  
	+ {\mathrm{cross\!-\!terms}} \nonumber\,.
\end{eqnarray}
Here 
$s=(p_a + p_b)^2$, $t=(p_a - p_c)^2$, $u=(p_a - p_d)^2$ 
are Mandelstam variables 
and $\kappa$ is an arbitrary renormalization scale. 

This paper crucially concerns itself 
with the far-infrared region of NGB pion momenta.
The imaginary part of $\ln\left(-\frac{s}{\kappa}\right)$ arises from the unitarity of the S-matrix and is related to a total cross-section.
The real part of $\ln\left(-\frac{s}{\kappa}\right)$ 
diverges in the far-infrared, 
and might have been important to \chiNL . We show that it is not! 
\footnote{
It clarifies things to regularize 
$\ln\left(-\frac{s}{\kappa}\right)\rightarrow \ln\left(-\frac{s_{IR}}{\kappa}\right)$ 
with $\vert s_{IR}\vert > 0$ in the Infra-Red.
}

Following  \cite{Kennedy19881}, 
we pack this non-analytic S-Matrix 
${\mathcal O}\left(\Lambda_{\chi SB}^0\right)$ term, 
and all other such non-analytic terms in the pure pion sector, 
into a non-analytic effective Lagrangian  
$L^{\pi; Symmetric}_{non-Analytic}$, 
which is also to be analyzed at tree-level.

\section{4-nucleon contact interactions in \SchiNL s}

\subsection{Boson-exchange-inspired vs. excited-nucleon-inspired 4-nucleon \\contact interactions}
We  wish to study the ground state expectation value of 
$L^{4-N;Symmetric}_{\chi PT}$.
Using \eqref{L4NSymmetricchiPT}
\begin{eqnarray}
&&\brao -L^{4-N;Symmetric}_{\chi PT} \keto = \\
 &&\quad\quad\quad\frac{1}{2f_\pi^2}\sum_{{\mathscr A}} \Big\{ \brao C^{T=0}_{\mathscr A}( \overline{N_c^\alpha}\gamma^{{\mathscr A}\alpha\beta} N_c^\beta)( \overline{N_e^\lambda}\gamma_{\mathscr A}^{\lambda\sigma} N_e^\sigma) \keto  \nonumber \\
&&\quad\quad\quad\quad\quad  + \frac{1}{4}\sum_B\brao C^{T=1}_{\mathscr A}( \overline{N_c^\alpha}\sigma^B_{cd}\gamma^{{\mathscr A}\alpha\beta} N_d^\beta)( \overline{N_e^\lambda}\sigma^B_{ef}\gamma_{\mathscr A}^{\lambda\sigma} N_f^\sigma)\keto \Big\}\,.\nonumber
\end{eqnarray}

Now introduce a complete set of states
\begin{eqnarray}
\one &=& \keto\brao + \sum_{\Psi \neq Static\chi NL}\big\vert \Psi \big>\big< \Psi \big\vert
\end{eqnarray}
and classify 4-nucleon \SchiNL~  interaction terms as either 
inspired by ``boson exchange" 
\begin{eqnarray}
\label{BosonExchangeAppendix_c}
&&-L_{Static \chi NL}^{BosonExchange} =\frac{1}{2f_\pi^2}\sum_{{\mathscr A}} \\
&&\quad \times \Big\{ C^{T=0}_{\mathscr A} \brao ( \overline{N_c^\alpha}\gamma^{{\mathscr A}\alpha\beta} N_c^\beta)
\keto\brao
( \overline{N_e^\lambda}\gamma_{\mathscr A}^{\lambda\sigma} N_e^\sigma) \keto  \nonumber \\
&&\quad \quad + \frac{1}{4} \sum_B C^{T=1}_{\mathscr A}\brao( \overline{N_c^\alpha}\sigma^B_{cd}\gamma^{{\mathscr A}\alpha\beta} N_d^\beta)
\keto \nonumber \\
&&\quad \quad  \times \brao ( \overline{N_e^\lambda}\sigma^B_{ef}\gamma_{\mathscr A}^{\lambda\sigma} N_f^\sigma)\keto \Big\}\nonumber
\end{eqnarray}
or ``excited-nucleon" inspired 
\begin{eqnarray}
\label{ExcitedNucleonAppendix_c}
&&-L_{Static \chi NL}^{ExcitedNucleon} =\frac{1}{2f_\pi^2}\sum_{{\mathscr A}} \\
&&\quad \times \Big\{ C^{T=0}_{\mathscr A} \brao ( \overline{N_c^\alpha}\gamma^{{\mathscr A}\alpha\beta} N_c^\beta)
( \overline{N_e^\lambda}\gamma_{\mathscr A}^{\lambda\sigma} N_e^\sigma) \keto  \nonumber \\
&&\quad \quad + \frac{1}{4} \sum_B C^{T=1}_{\mathscr A}\brao( \overline{N_c^\alpha}\sigma^B_{cd}\gamma^{{\mathscr A}\alpha\beta} N_d^\beta)
 ( \overline{N_e^\lambda}\sigma^B_{ef}\gamma_{\mathscr A}^{\lambda\sigma} N_f^\sigma)\keto \Big\}\nonumber
\end{eqnarray}

A useful theorem is
\begin{eqnarray}
\label{UsefulTheorem}
&&  \frac{1}{4}\brao ( \overline{N_c^\alpha}\gamma^{{\mathscr A}\alpha\beta} N_c^\beta)
\keto\brao
( \overline{N_e^\lambda}\gamma_{\mathscr A}^{\lambda\sigma} N_e^\sigma) \keto  \nonumber \\
&&+   \brao( \overline{N_c^\alpha} t_{3;cd}\gamma^{{\mathscr A}\alpha\beta} N_d^\beta)
\keto 
\brao ( \overline{N_e^\lambda}t_{3;ef}\gamma_{\mathscr A}^{\lambda\sigma} N_f^\sigma)\keto \nonumber \\
&&  = \half \brao ( \overline{p_c^\alpha}\gamma^{{\mathscr A}\alpha\beta} p_c^\beta)
\keto\brao
( \overline{p_e^\lambda}\gamma_{\mathscr A}^{\lambda\sigma} p_e^\sigma) \keto  \nonumber \\
&&  +\half \brao ( \overline{n_c^\alpha}\gamma^{{\mathscr A}\alpha\beta} n_c^\beta)
\keto\brao
( \overline{n_e^\lambda}\gamma_{\mathscr A}^{\lambda\sigma} n_e^\sigma) \keto  \nonumber
\end{eqnarray}

Going forward, we will use the notation $\Big> \equiv \keto$ and $\Big< \equiv \brao$ in this Appendix.

\subsection{Contact-interactions that mimic hadronic boson-exchange}
Taking expectation values inside the \SchiNL, 
\begin{eqnarray}
\label{BosonExchangeAppendix_c2}
&&-L_{Static \chi NL}^{BosonExchange} \simeq\half \frac{1}{f\pi} \\
&& \qquad \times \Big[ C^{T=0}_{S} \Big< \overline{N_c^\alpha} N_c^\alpha\Big>\Big< \overline{N_e^\lambda} N_e^\lambda \Big> 
\nonumber \\
&&\qquad +C^{T=0}_{V} \Big< \overline{N_c^\alpha}\gamma^{0;\alpha\beta} N_c^\beta\Big>\Big< \overline{N_e^\lambda}\gamma_{0}^{\lambda\sigma} N_e^\sigma\Big>  \nonumber \\
&& \qquad+C^{T=1}_{S}\Big\{ \frac{1}{4}\Big< \overline{N_c^\alpha} N_c^\alpha\Big>
 \Big< \overline{N_e^\lambda} N_e^\lambda\Big>+\Big< \overline{N_c^\alpha} {t}_{3;cd} N_d^\alpha\Big>
\Big< \overline{N_e^\lambda}{t}_{3;ef} N_f^\lambda\Big> \Big\}\nonumber \\
&&\qquad+C^{T=1}_{V}\Big\{ \frac{1}{4}\Big< \overline{N_c^\alpha}\gamma^{0;\alpha\beta}N_c^\beta\Big>
 \Big<\overline{N_e^\lambda}\gamma_{0}^{\lambda\sigma} N_e^\sigma\Big> +\Big< \overline{N_c^\alpha} {t}_{3;cd}\gamma^{0;\alpha\beta} N_d^\beta\Big>
 \Big< \overline{N_e^\lambda}{t}_{3;ef}\gamma_{0}^{\lambda\sigma} N_f^\sigma\Big> \Big\} \Big]\,.\nonumber
\end{eqnarray}

The factorization in $L_{Static \chi NL}^{BosonExchange}$, 
 and its name, are inspired by a simple picture 
of forces carried by heavy hadronic-boson exchange;
i.e. we have integrated out the auxiliary fields:
\begin{itemize}
\item  Lorentz-scalar isoscalar $\sigma$, with chiral coefficient $C^{T=0}_{S}$~;
\item  Lorentz-vector isoscalar $\omega_\mu$ with chiral coefficient $C^{T=0}_{V}$~;
\item  Lorentz-scalar isovector ${\vec \delta}$, with chiral coefficient $C^{T=1}_{S}$~~;
\item  Lorentz-vector isovector ${\vec \rho}_\mu$, with chiral coefficient $C^{T=1}_{V}$.
\end{itemize}

To order $\Lambda^0_{\chi SB}$, 
the only 4-nucleon contact terms allowed 
by local $SU(2) \chi PT$ symmetry 
are  exhibited in \eqref{BosonExchangeAppendix_c} 
(i.e. \eqref{BosonExchangeAppendix_c2}) 
and \eqref{ExcitedNucleonAppendix_c}.
Note that isospin operators $\vec{t} = \half\vec{\sigma}_{Pauli}$  have appeared.
However, quantum-loop power counting 
requires inclusion of nucleon Lorentz-spinor-interchange interactions,
in order to enforce anti-symmetrization of fermion wavefunctions. 
These are the same magnitude, $(\Lambda{\chi SB})^0$, as direct interactions.
The empirical nuclear models 
of Manakos and Mannel \cite{Manakos1988223, Manakos1989481} 
were specifically built to include such spinor-interchange terms. 

Explicit inclusion of spinor-interchange terms yields a great technical advantage for the liquid approximation: it allows us to treat \SchiNL s  in Hartree-Fock approximation, i.e. including fermion wavefunction anti-symmetrization, rather than in less-accurate Hartree approximation. 

Because of normal-ordering, 
such point-coupling contact spinor-interchange terms 
don't appear in the analysis of the deuteron \cite{Weinberg1965672, vanKolck19942932}, 
which has only 1 proton and 1 neutron.

\subsection{Contact-interactions, 
including spinor-interchange terms enforcing \\ effective anti-symmetrization of fermion wavefunctions in the \\Hartree-Fock approximation}
\label{ContactSpinorInterchange}

In this section, we write an effective \SchiNL~  Lagrangian in terms of the 10 independent chiral coefficients
${C^{T=0}_{S}}$, $ {C^{T=0}_{V}}$, $ {C^{T=0}_{T}}$,
 $ {C^{T=0}_{A}}$,$ {C^{T=0}_{P}}$ and
${C^{T=1}_{S}}$, $ {C^{T=1}_{V}}$, $ {C^{T=1}_{T}}$,
 $ {C^{T=1}_{A}}$, $ {C^{T=1}_{P}}$ governing 4-nucleon contact interactions.

For pedagogical simplicity, 
we first focus on the ``boson-exchange-inspired" terms, 
with power-counting contact-interactions of order $(\Lambda{\chi SB})^0$.
``Direct" terms depend only on 
${C^{T=0}_{S}}$, ${C^{T=0}_{V}}$,${C^{T=1}_{S}}$, 	and ${C^{T=1}_{V}}$, 
because isoscalar (${C^{T=0}_{T}}$, ${C^{T=0}_{A}}$, and ${C^{T=0}_{P}}$) 
and isovector (${C^{T=1}_{T}}$, ${C^{T=1}_{A}}$, ${C^{T=1}_{P}}$) 
vanish when evaluated in the liquid.
``Spinor-interchange" terms depend all 10 coefficients after Fierz rearrangement.
Such terms do not appear in the \chiPTtw~ analysis 
of the deuteron ground state,
because it only has 1 proton and 1 neutron.
The combination of direct and spinor-interchange terms 
(which we refer to below as  ``Total'')
depend on all 10 coefficients. 

Because of the inclusion of spinor-interchange terms,
Hartree treatment of the resultant \\ \SchiNL~ Lagrangian 
is equivalent to Hartree-Fock treatment of the liquid. 
When building the semi-classical liquid quantum state, 
this enforces the anti-symmetrization of the fermion wavefunctions.
A crucial observation is that the resultant liquid 
depends on only four independent chiral coefficients. 
${C^{S}_{200}}$, ${C^{V}_{200}}$, ${\overline {C^{S}_{200}}}$, 
and ${\overline{C^{V}_{200}}}$.
These provide sufficient free parameters to balance 
the scalar repulsive force carried by 
$C^{S}_{200}$ and ${\overline {C^{S}_{200}}}$ 
against the vector repulsive force carried by 
$C^{V}_{200}$ and ${\overline {C^{V}_{200}}}$ 
when fitting to the experimentally observed structure of ground-state nuclei. 
This is the case for our Non-topological Soliton nuclear model, 
where $C^{S}_{200}-\half {\overline {C^{S}_{200}}}<0$ 
and $C^{V}_{200}-\half {\overline {C^{S}_{200}}}>0$, 
and we conjecture it to persist in
Density Functional and Skyrme nuclear models.

Motivated by the empirical success 
of Non-topological Soliton, Density Functional and Skyrme nuclear models, 
we also conjecture that 
excited-nucleon-inspired contact-interaction terms are small, 
and that the simple picture of  scalar attraction 
balanced against vector repulsion persists when including them. 
But such analysis is beyond the scope of this paper.

\subsubsection{Lorentz Vector (V) and Axial-vector (A) forces}

\begin{eqnarray}
\label{VectorBosonExchange}
\Big<L^{4-N;V,A} \Big>&\equiv&L_{Static \chi NL}^{V,A} \nonumber\\
-L_{Static \chi NL}^{V,A} &=&
	\frac{1}{2f_\pi^2}\sum_{{\mathscr A}=V,A} \Big\{ C^{T=0}_{\mathscr A} \Big< ( \overline{N_c^\alpha}\gamma^{{\mathscr A}\alpha\beta} N_c^\beta)
\Big>\Big<
( \overline{N_e^\lambda}\gamma_{\mathscr A}^{\lambda\sigma} N_e^\sigma) \Big> \nonumber \\
&&\quad \quad + \frac{1}{4} \sum_B C^{T=1}_{\mathscr A}\Big< ( \overline{N_c^\alpha}\sigma^B_{cd}\gamma^{{\mathscr A}\alpha\beta} N_d^\beta)
\Big> \Big<( \overline{N_e^\lambda}\sigma^B_{ef}\gamma_{\mathscr A}^{\lambda\sigma} N_f^\sigma)\Big> \Big\}\nonumber \\
&&\quad -L_{Static \chi NL;ExcitedNucleon}^{V,A}  \\
-L_{Static \chi NL;ExcitedNucleon}^{V,A} &=&
\frac{1}{2f_\pi^2}\sum_{{\mathscr A}=V,A} \Big\{ C^{T=0}_{\mathscr A} \Big< ( \overline{N_c^\alpha}\gamma^{{\mathscr A}\alpha\beta} N_c^\beta)
( \overline{N_e^\lambda}\gamma_{\mathscr A}^{\lambda\sigma} N_e^\sigma) \Big> \nonumber  \\
&&\quad \quad+ \frac{1}{4} \sum_B C^{T=1}_{\mathscr A}\Big< ( \overline{N_c^\alpha}\sigma^B_{cd}\gamma^{{\mathscr A}\alpha\beta} N_d^\beta)
( \overline{N_e^\lambda}\sigma^B_{ef}\gamma_{\mathscr A}^{\lambda\sigma} N_f^\sigma)\Big> \Big\}\nonumber
\end{eqnarray}

We have
\begin{eqnarray}
\label{RewriteVectorAxialvector}
-L_{Static \chi NL}^{V,A} &=&\frac{1}{2f^2_\pi} \sum_{{\mathscr A}=V,A}\Big[C^{T=0}_{\mathscr A} \Big\{ 2 \Big< ( \overline{p_c^\alpha}\gamma^{{\mathscr A}\alpha\beta} p_c^\beta)
\Big>\Big<
( \overline{n_e^\lambda}\gamma_{\mathscr A}^{\lambda\sigma} n_e^\sigma) \Big>  \Big\} \nonumber \\
& +&  \Big[ C^{T=0}_{\mathscr A} +\half C^{T=1}_{\mathscr A} \Big] \Big\{  \Big< ( \overline{p_c^\alpha}\gamma^{{\mathscr A}\alpha\beta} p_c^\beta)
\Big>\Big<
( \overline{p_e^\lambda}\gamma_{\mathscr A}^{\lambda\sigma} p_e^\sigma) \Big>  \\
& &  \quad \quad + \Big< ( \overline{n_c^\alpha}\gamma^{{\mathscr A}\alpha\beta} n_c^\beta)
\Big>\Big<
( \overline{n_e^\lambda}\gamma_{\mathscr A}^{\lambda\sigma} n_e^\sigma) \Big> \Big\} \Big] \nonumber \\
&-&L_{Static \chi NL;ExcitedNucleon}^{V,A} \nonumber
\end{eqnarray}

\paragraph{Direct terms:}
The properties of \SchiNL s vastly simplify this expression
\begin{eqnarray}
\label{SimplifyDirectVectorAxialvector1}
-L_{Static \chi NL;Direct}^{V,A} &=&\frac{1}{2f^2_\pi} C^{T=0}_{V} \Big\{ 2 \Big< {p^{\dagger}}p\Big> \Big< {n^{\dagger}}n \Big>  \Big\}\nonumber\\
& +& \frac{1}{2f^2_\pi} \Big[ C^{T=0}_{V} +\half C^{T=1}_{V}  \Big] \Big\{  \Big< {p^{\dagger}}p \Big>
 \Big< {p^\dagger}p \Big> +\Big< {n^{\dagger}}n \Big>
 \Big< {n^\dagger}n \Big> \Big\} \nonumber \\
&-&L_{Static \chi NL;ExcitedNucleon;Direct}^{V,A}
\end{eqnarray}
with simplified notation
$  \Big< {p_c^{\alpha\dagger}}p_c^\alpha \Big> \Big< {n_e^{\lambda;\dagger}}n_e^\lambda \Big> \equiv  \Big< {p^{\dagger}}p\Big> \Big< {n^{\dagger}}n \Big> $.

\paragraph{Spinor-interchange terms:}
After interchanging the appropriate spinors, normal ordering creation and annihilation operators, and Fierz re-arrangement, spinor-interchange contributions depend on $C^{T=0}_{V}, C^{T=0}_{A},C^{T=1}_{V}, C^{T=1}_{A}$. 
\begin{eqnarray}
\label{SimplifyDirectVectorAxialvector2}
&&-L_{Static \chi NL;SpinorInterchange}^{V,A} = \nonumber\\
&&\qquad\qquad \frac{1}{2f^2_\pi} 
\Big[ -\Big(C^{T=0}_{V} +\half C^{T=1}_{V} \Big) + \Big(C^{T=0}_{A} +\half C^{T=1}_{A} \Big)  \Big] \nonumber \\
&&\qquad \qquad \times \Big\{  \Big< {p_L^{\dagger}}p_L \Big>
 \Big< {p_L^\dagger}p_L \Big> + \Big< {p_R^{\dagger}}p_R \Big>\Big< {p_R^\dagger}p_R \Big> \nonumber\\
 &&\qquad\qquad\qquad\qquad + \Big< {n_L^\dagger}n_L \Big>\Big< {n_L^{\dagger}}n_L \Big>+\Big< {n_R^{\dagger}}n_R \Big>
 \Big< {n_R^\dagger}n_R \Big>\Big\} \nonumber \\
&&\qquad \qquad -L_{Static \chi NL;ExcitedNucleon;SpinorInterchange}^{V,A}
\end{eqnarray}
where we have divided $p=p_L+p_R$ and $n=n_L+n_R$ into left-handed and right-handed spinors.

\paragraph{Total direct and spinor-interchange terms:}
\begin{eqnarray}
\label{TotalVectorAxialvector}
&&-L_{Static \chi NL;Total}^{V,A} =\frac{1}{2f^2_\pi} C^{T=0}_{V} \Big\{ 2 \Big< {p^{\dagger}}p\Big> \Big< {n^{\dagger}}n \Big>  \Big\}\\
&&\qquad \qquad+ \frac{1}{2f^2_\pi} 
\Big[ C^{T=0}_{V} +\half C^{T=1}_{V}   \Big] \Big\{ 2\Big< {p_L^{\dagger}}p_L \Big>
 \Big< {p_R^\dagger}p_R \Big> +2 \Big< {n_L^\dagger}n_L \Big>\Big< {n_R^{\dagger}}n_R \Big>\Big\}\nonumber \\
 &&\qquad \qquad+ \frac{1}{2f^2_\pi} 
\Big[  C^{T=0}_{A} +\half C^{T=1}_{A}\Big] 
	\Big\{  \Big< {p_L^{\dagger}}p_L \Big>
 \Big< {p_L^\dagger}p_L \Big> + \Big< {p_R^{\dagger}}p_R \Big>\Big< {p_R^\dagger}p_R \Big> \nonumber\\
&&\qquad\qquad\qquad\qquad\qquad\qquad + \Big< {n_L^\dagger}n_L \Big>\Big< {n_L^{\dagger}}n_L \Big>+\Big< {n_R^{\dagger}}n_R \Big>
 \Big< {n_R^\dagger}n_R \Big>\Big\} \nonumber \\
&&\qquad \qquad -L_{Static \chi NL;ExcitedNucleon;Total}^{V,A}\,.\nonumber
\end{eqnarray}

The reader should note the cancellation of the term
\begin{eqnarray}
\label{VectorCancellation}
\frac{1}{2f_\pi^2}\Big[  C^{T=0}_{V} +\half C^{T=1}_{V}   \Big] 
&&\Big\{ \Big< {p_L^{\dagger}}p_L \Big>\Big< {p_L^\dagger}p_L \Big> 
 + \Big< {p_R^{\dagger}}p_R \Big>\Big< {p_R^\dagger}p_R \Big> \nonumber\\
&&\qquad + \Big< {n_L^\dagger}n_L \Big>\Big< {n_L^{\dagger}}n_L \Big>+\Big< {n_R^{\dagger}}n_R \Big>
 \Big< {n_R^\dagger}n_R \Big>\Big\}\,, \nonumber 
\end{eqnarray}
showing that vector-boson exchange 
cannot carry forces between same-handed protons, 
or between same-handed neutrons.

Significant simplification follows because 
\SchiNL s are defined to have equal left-handed and right-handed densities
\begin{eqnarray}
 \Big< p^\dagger_L p_L\Big>&=&\Big< p^\dagger_R p_R\Big>=\half\Big< p^\dagger p\Big> \\
 \Big< n^\dagger_L n_L\Big>&=&\Big< n^\dagger_R n_R\Big>=\half\Big< n^\dagger n\Big>\,.\nonumber
\end{eqnarray}
so that the contribution of (\ref{TotalVectorAxialvector}) 
to the Lorentz-spinor-interchange Lagrangian is
\begin{eqnarray}
\label{VectorAxialvectorChiralCoefficients}
	&&-L_{Static \chi NL;Total}^{V,A} =\frac{1}{2f^2_\pi} C^{V}_{200} \Big\{  \Big< {N^{\dagger}}N \Big> \Big< {N^{\dagger}}N \Big>  \Big\}\\
	 &&\qquad \qquad - \frac{1}{4f^2_\pi} {\overline {C^{V}_{200}}}  \Big\{ \Big< {N^{\dagger}}N \Big> \Big< {N^{\dagger}}N \Big> 
	+ 4\Big< {N^{\dagger}}t_3 N \Big> \Big< {N^{\dagger}}t_3 N \Big> 
	 \Big\} \nonumber \\
	&&\qquad \qquad -L_{Static \chi NL;ExcitedNucleon;Total}^{V,A} \nonumber
\end{eqnarray}
with 
\begin{eqnarray}
\label{VectorNuclearCoefficients}
C^{V}_{200}&=&C^{T=0}_{V} \\
-{\overline {C^{V}_{200}}} &=& \half \Big[ -C^{T=0}_{V}+C^{T=0}_{A} +\half C^{T=1}_{V}+\half C^{T=1}_{A}\Big] \nonumber
\end{eqnarray}

The crucial observation is that 
(\ref{VectorAxialvectorChiralCoefficients}, \ref{VectorNuclearCoefficients})
depend on just {\em{two}} independent chiral coefficients, 
$C^{V}_{200}$ and ${\overline {C^{V}_{200}}}$, 
instead of four, 
while still providing sufficient free parameters 
to fit the vector repulsive force 
(i.e. within Non-topological Soliton, Density Functional 
and Skyrme nuclear models)
up to power-counting order $(\Lambda{\chi SB})^0$, 
to the experimentally observed structure of ground-state nuclei.

\subsubsection{Lorentz Scalar (S), Tensor (T) and Pseudo-scalar (P) forces}

\begin{eqnarray}
\label{ScalarBosonExchange}
	\Big<L^{4-N;ScalarTensorPseudoscalar} \Big>&\equiv&L_{Static \chi NL}^{ScalarTensorPseudoscalar} \\
	-L_{Static \chi NL}^{ScalarTensorPseudoscalar} 
	&=&\frac{1}{2f_\pi^2}\sum_{{\mathscr A}=S,T,P} 
		\Big\{ C^{T=0}_{\mathscr A} \Big< ( \overline{N_c^\alpha}\gamma^{{\mathscr A}\alpha\beta} N_c^\beta)
	\Big>\Big<
	( \overline{N_e^\lambda}\gamma_{\mathscr A}^{\lambda\sigma} N_e^\sigma) \Big>  \nonumber \\
	&&\quad \quad + \frac{1}{4} \sum_B C^{T=1}_{\mathscr A}\Big< ( \overline{N_c^\alpha}\sigma^B_{cd}\gamma^{{\mathscr A}\alpha\beta} N_d^\beta)
	\Big> \Big<( \overline{N_e^\lambda}\sigma^B_{ef}\gamma_{\mathscr A}^{\lambda\sigma} N_f^\sigma)\Big> \Big\}\nonumber \\
	&-&L_{Static \chi NL;ExcitedNucleon}^{ScalarTensorPseudoscalar} \nonumber \\
	-L_{Static \chi NL;ExcitedNucleon}^{ScalarTensorPseudoscalar} 
	&=&\frac{1}{2f_\pi^2}\sum_{{\mathscr A}=S,T,P} 
		\Big\{ C^{T=0}_{\mathscr A} \Big< ( \overline{N_c^\alpha}\gamma^{{\mathscr A}\alpha\beta} N_c^\beta)
	( \overline{N_e^\lambda}\gamma_{\mathscr A}^{\lambda\sigma} N_e^\sigma) \Big> \nonumber  \\
	&&\quad \quad+ \frac{1}{4} \sum_B C^{T=1}_{\mathscr A}\Big< ( \overline{N_c^\alpha}\sigma^B_{cd}\gamma^{{\mathscr A}\alpha\beta} N_d^\beta)
	( \overline{N_e^\lambda}\sigma^B_{ef}\gamma_{\mathscr A}^{\lambda\sigma} N_f^\sigma)\Big> \Big\}\Big]\nonumber
\end{eqnarray}

We have
\begin{eqnarray}
\label{RewriteScalarTensorPseudoscalar}
-L_{Static \chi NL}^{ScalarTensorPseudoscalar} &=&\frac{1}{2f^2_\pi} \sum_{{\mathscr A}=S,T,P} \Big[C^{T=0}_{\mathscr A} \Big\{ 2 \Big< ( \overline{p_c^\alpha}\gamma^{{\mathscr A}\alpha\beta} p_c^\beta)
\Big>\Big<
( \overline{n_e^\lambda}\gamma_{\mathscr A}^{\lambda\sigma} n_e^\sigma) \Big>  \Big\} \nonumber \\
& +& \Big[ C^{T=0}_{\mathscr A} +\half C^{T=1}_{\mathscr A} \Big] \Big\{  \Big< ( \overline{p_c^\alpha}\gamma^{{\mathscr A}\alpha\beta} p_c^\beta)
\Big>\Big<
( \overline{p_e^\lambda}\gamma_{\mathscr A}^{\lambda\sigma} p_e^\sigma) \Big>  \\
& &  \qquad \qquad + \Big< ( \overline{n_c^\alpha}\gamma^{{\mathscr A}\alpha\beta} n_c^\beta)
\Big>\Big<
( \overline{n_e^\lambda}\gamma_{\mathscr A}^{\lambda\sigma} n_e^\sigma) \Big> \Big\}  \nonumber \\
&-&L_{Static \chi NL;ExcitedNucleon}^{ScalarTensorPseudoscalar} \nonumber
\end{eqnarray}

\paragraph{Direct terms:}
The properties of \SchiNL s give
\begin{eqnarray}
\label{SimplifyDirectScalarTensorPseudoscalar1}
-L_{Static \chi NL;Direct}^{ScalarTensorPseudoscalar} &=&\frac{1}{2f^2_\pi}  C^{T=0}_{S}  \Big< {\overline N}N\Big> \Big< {\overline N}N \Big>  \\
& +&\frac{1}{2f^2_\pi} \big( \half C^{T=1}_{S} \big) \Big\{  \Big< {\overline p}p \Big>
 \Big< {\overline p}p \Big> +\Big< {\overline n}n \Big>
 \Big< {\overline n}n \Big> \Big\}  \nonumber \\
&-&L_{Static \chi NL;ExcitedNucleon;Direct}^{ScalarTensorPseudoscalar}\nonumber
\end{eqnarray}

\paragraph{Spinor-interchange terms:}
Spinor-interchange contributions depend on 6 chiral coefficients: isoscalars $C^{T=0}_{S}$, $C^{T=0}_{T}$, $C^{T=0}_{P}$ 
and isovectors $C^{T=1}_{S}$, $C^{T=1}_{T}$, $C^{T=1}_{P}$. 

\begin{eqnarray}
\label{SimplifyDirectScalarTensorPseudoscalar2}
&&-L_{Static \chi NL;SpinorInterchange}^{ScalarTensorPseudoscalar} = \\
&&\qquad\qquad \frac{1}{2f^2_\pi} 
\Big[ \half \Big(C^{T=0}_{S} +\half C^{T=1}_{S} \Big) + 3 \Big(C^{T=0}_{T} +\half C^{T=1}_{T} \Big) + \half \Big(C^{T=0}_{P} +\half C^{T=1}_{P} \Big)  \Big] \nonumber \\
&&\qquad \qquad \times \Big\{ 
\Big< {\overline p_L}p_R \Big> \Big< {\overline p_L}p_R \Big>
 + \Big< {\overline p_R}p_L \Big> \Big< {\overline p_R}p_L \Big>
  +\Big< {\overline n_L}n_R \Big> \Big< {\overline n_L}n_R \Big>
 + \Big< {\overline n_R}n_L \Big> \Big< {\overline n_R}n_L \Big> 
\Big\} \nonumber \\
&&\qquad \qquad -L_{Static \chi NL;ExcitedNucleon;SpinorInterchange}^{ScalarTensorPseudoscalar}\nonumber
\end{eqnarray}

\paragraph{Total direct and spinor-interchange terms:}

As above, the fact that \SchiNL s are defined 
to have equal left-handed and right-handed scalar densities
simplifies
the total direct and spinor-interchange contribution:
\begin{eqnarray}
\label{ScalarTensorPseudoscalarChiralCoefficients}
&&-L_{Static \chi NL;Total}^{ScalarTensorPseudoscalar} =\frac{1}{2f^2_\pi} C^{S}_{200} \Big\{  \Big< {\overline N}N \Big> \Big< {\overline N}N \Big>  \Big\}\nonumber\\
 &&\qquad \qquad - \frac{1}{4f^2_\pi} {\overline {C^{S}_{200}}}  \Big\{ \Big< {\overline N}N \Big> \Big< {\overline N}N \Big> 
+ 4\Big< {\overline N}t_3 N \Big> \Big< {\overline N}t_3 N \Big> 
 \Big\} \nonumber \\
&&\qquad \qquad -L_{Static \chi NL;ExcitedNucleon;Total}^{ScalarTensorPseudoscalar}
\end{eqnarray}
with 
\begin{eqnarray}
\label{ScalarNuclearCoefficients}
C^{S}_{200}&=&C^{T=0}_{S} \\
-{\overline {C^{S}_{200}}} &=& \half \Big[ \half C^{T=0}_{S}+ \frac{5}{4} C^{T=1}_{S}+3\Big( C^{T=0}_{T} +\half C^{T=1}_{T}\Big) +\half \Big( C^{T=0}_{P}+ \half C^{T=1}_{P} \Big) \Big] \nonumber
\end{eqnarray}

Once again we find that 
(\ref{ScalarTensorPseudoscalarChiralCoefficients}, 
	\ref{ScalarNuclearCoefficients})
depend on just {\em{two}} independent chiral coefficients, 
$C^{S}_{200}$ and ${\overline {C^{S}_{200}}}$, instead of six, 
while still providing sufficient free parameters 
to fit the scalar attractive force 
(i.e. within Non-topological Soliton, Density Functional 
and Skyrme nuclear models) 
up to power-counting order $(\Lambda{\chi SB})^0$, 
to the experimentally observed structure of ground-state nuclei.

\section{Nucleon bi-linears and nuclear currents in \\ \SchiNL }

The structure of \SchiNL~  suppresses various nucleon bi-linears:

\begin{itemize}

\item Vectors' space-components: because it is a 3-vector, parity odd and stationary 

                \begin{eqnarray}
                \brao( \overline{N_c^\alpha}{\vec \gamma}^{\alpha\beta} N_c^\beta)\keto \sim  \brao {\vec k}\keto \simeq 0 \quad \quad
                \end{eqnarray}

\item Tensors: because the local expectation value of the nuclear spin ${\vec s}=\half {\vec \sigma}\simeq 0$ 

\begin{enumerate}

\item $\sigma^{0j}$:
            \begin{eqnarray}
 &&\brao( \overline{N_c^\alpha}\sigma^{0j;\alpha\beta} N_c^\beta)\keto  \nonumber \\
&& \quad \quad =\brao( \overline{N_L}\sigma^{0j} N_R)\keto +\brao( \overline{N_R}\sigma^{0j} N_L)
\keto  \nonumber \\
&& \quad \quad = 2\brao( \overline{N_L} 
{ \begin{aligned}
 \label{pGoctetrepn2}
    \left [
    \begin{tabular}{ccc} 
    $0$ & ${\vec s}_j$ \\ 
    ${\vec s}_j$ & $0$      \\
    \end{tabular} \right ]
  \end{aligned}}
 N_R)\keto \nonumber \\
 && \quad \quad +2\brao( \overline{N_R}
 { \begin{aligned}
 \label{pGoctetrepn3}
    \left [
    \begin{tabular}{ccc} 
    $0$ & ${\vec s}_j$ \\ 
    ${\vec s}_j$ & $0$      \\
    \end{tabular} \right ]
  \end{aligned}}
  N_L)
\keto  \nonumber \\
&& \quad \quad \simeq 0
                \end{eqnarray}
                
\item $\sigma^{ij}$:
            \begin{eqnarray}
 &&\brao( \overline{N_c^\alpha}\sigma^{ij;\alpha\beta} N_c^\beta)\keto  \nonumber \\
&& \quad \quad =\brao( \overline{N_L}\sigma^{ij} N_R)\keto +\brao( \overline{N_R}\sigma^{ij} N_L)
\keto  \nonumber \\
&& \quad \quad = -2i\epsilon_{ijk}\brao( \overline{N_L} {\vec s}_k N_R)\keto \nonumber \\
&& \quad \quad \quad -2i\epsilon_{ijk}\brao( \overline{N_R} {\vec s}_k N_L)\keto\nonumber \\
 && \quad \quad \simeq 0
                \end{eqnarray}

\end{enumerate}

\item Axial-vectors: because ${p_L},{p_R}$ are equally represented in \SchiNL, as are ${n_L},{n_R}$              
  \begin{eqnarray}
 &&\brao( \overline{N_c^\alpha}\gamma^{A;\alpha\beta} N_c^\beta)
\keto  \nonumber \\
&& \quad \quad =\brao( \overline{N_L}\gamma^\mu \gamma^5 N_L)\keto +\brao( \overline{N_R}\gamma^\mu \gamma^5 N_R)
\keto  \nonumber \\
&& \quad \quad = -\brao( \overline{N_L}\gamma^\mu N_L)\keto +\brao( \overline{N_R} \gamma^\mu N_R)
\keto  \nonumber \\
&& \quad \quad \simeq 0
                \end{eqnarray}

\item Pseudo-scalars: because  \SchiNL~  are of even parity
                \begin{eqnarray}
 &&\brao( \overline{N_c^\alpha}\gamma^{P;\alpha\beta} N_c^\beta)
\keto  \nonumber \\
&& \quad \quad =\brao( \overline{N_R}\gamma^5 N_L)\keto +\brao( \overline{N_L}\gamma^5 N_R)
\keto  \nonumber \\
&& \quad \quad = -\brao( \overline{N_R} N_L)\keto +\brao( \overline{N_L} N_R)
\keto  \nonumber \\
&& \quad \quad \simeq 0
                \end{eqnarray}

\end{itemize}

Therefore, various Lorentz and isospin representations are suppressed in \SchiNL s. In summary: Isoscalars
\begin{eqnarray}
&&  \brao( \overline{N_c^\alpha} N_c^\alpha)
\keto \neq 0\nonumber \\
&&  \brao( \overline{N_c^\alpha}\gamma^{0;\alpha\beta} N_c^\beta)
\keto \neq 0\nonumber \\
&&  \brao( \overline{N_c^\alpha}{\vec \gamma}^{\alpha\beta} N_c^\beta)
\keto \simeq 0\nonumber \\
&&  \brao( \overline{N_c^\alpha}\gamma^{T;\alpha\beta} N_c^\beta)
\keto \simeq 0\nonumber \\
&&  \brao( \overline{N_c^\alpha}\gamma^{A;\alpha\beta} N_c^\beta)
\keto \simeq 0\nonumber \\
&&  \brao( \overline{N_c^\alpha}\gamma^{P;\alpha\beta} N_c^\beta)
\keto \simeq 0
\end{eqnarray}
and Isovectors
\begin{eqnarray}
 && \brao( \overline{N_c^\alpha} {t}^\pm_{cd}\gamma^{{\mathscr A}\alpha\beta} N_d^\beta)
\keto = 0
\nonumber \\
 && \brao( \overline{N_c^\alpha} {t}^3_{cd}N_d^\alpha)
\keto \neq0
\nonumber \\
 && \brao( \overline{N_c^\alpha} {t}^3_{cd}\gamma^{0;\alpha\beta} N_d^\beta)
\keto \neq 0
\nonumber \\
 && \brao( \overline{N_c^\alpha} {t}^3_{cd}{\vec \gamma}^{\alpha\beta} N_d^\beta)
\keto \simeq 0
\nonumber \\
 && \brao( \overline{N_c^\alpha} {t}^3_{cd}\gamma^{T\alpha\beta} N_d^\beta)
\keto \simeq 0
\nonumber \\
 && \brao( \overline{N_c^\alpha} {t}^3_{cd}\gamma^{A\alpha\beta} N_d^\beta)
\keto \simeq 0
\nonumber \\
 && \brao( \overline{N_c^\alpha} {t}^3_{cd}\gamma^{P\alpha\beta} N_d^\beta)
\keto \simeq 0
\end{eqnarray}

Now form the nuclear currents
\begin{eqnarray}
  \label{nucleonbilinears}
  J^\mu_k &=& \overline{N}\gamma^\mu t_k N \quad k = 1, 2, 3\nonumber \\
  J^\mu_\pm &=& J^\mu_1 \pm iJ^\mu_2 = \left \{ \begin{tabular}{c} $\overline{p}\gamma^\mu n$ \\ $\overline{n}\gamma^\mu p$ \end{tabular} \right \} \nonumber \\
  J^\mu_3 &=& \frac{1}{2}\left ( \overline{p}\gamma^\mu p - \overline{n}\gamma^\mu n \right ) \nonumber \\
  J^\mu_8 &=& \frac{\sqrt{3}}{2}\left ( \overline{p}\gamma^\mu p + \overline{n}\gamma^\mu n \right )\nonumber \\
   J^\mu_{QED} &=& \frac{1}{\sqrt{3}}  J^\mu_8 + J^\mu_3 = \overline{p}\gamma^\mu p \nonumber \\
     J^\mu_{Baryon} &=& \frac{2}{\sqrt{3}}  J^\mu_8  = \overline{p}\gamma^\mu p +  \overline{n}\gamma^\mu n\nonumber \\
  J^{5\mu}_k &=& \overline{N}\gamma^\mu\gamma^5 t_k N \quad
  k = 1, 2, 3 \nonumber \\
  J^{5\mu}_\pm &=& J^{5\mu}_1 \pm iJ^{5\mu}_2 = \left \{ \begin{tabular}{c} $\overline{p}\gamma^\mu \gamma^5 n$ \\ $\overline{n}\gamma^\mu  \gamma^5 p$ \end{tabular} \right \}  \nonumber \\
  J^{5\mu}_{3} &=& \frac{1}{2}\left ( \overline{p}\gamma^\mu \gamma^5 p - \overline{n}\gamma^\mu \gamma^5 n \right )  \nonumber \\
  J^{5\mu}_{8} &=& \frac{\sqrt{3}}{2}\left ( \overline{p}\gamma^\mu \gamma^5 p + \overline{n}\gamma^\mu \gamma^5 n \right )
\end{eqnarray}

$SU(2)_L\times SU(2)_R$ nucleon currents within \SchiNL~  are obedient to its symmetries
 \begin{eqnarray}
 \label{VanishingCurrentsAppendix}
  \brao {J}^{0}_3 \keto &\neq& 0 ;\quad  \brao \partial_\mu{J}_{3}^\mu \keto \simeq 0  \nonumber \\
  \brao {J}^{\mu}_\pm \keto &=&  0; \quad  \brao \partial_\mu{J}_{\pm}^\mu \keto = 0 \nonumber \\
 \brao {J}^{\mu,5}_3 \keto &\simeq& 0 ;\quad  \brao \partial_\mu{J}^{\mu,5}_3 \keto \simeq 0 \nonumber \\
 \brao {J}^{\mu, 5}_\pm \keto &=& 0; \quad  \brao \partial_\mu{J}^{\mu, 5}_\pm \keto =0\nonumber \\
   \brao {J}^{0}_8 \keto &\neq& 0 ;\quad  \brao \partial_\mu{J}_{8}^\mu \keto \simeq 0  \nonumber \\
  \brao {J}^{0}_{QED} \keto &\neq& 0 ;\quad  \brao \partial_\mu{J}_{QED}^\mu \keto \simeq 0 \nonumber \\
   \brao {J}^{0}_{Baryon} \keto &\neq& 0 ;\quad  \brao \partial_\mu{J}_{Baryon}^\mu \keto \simeq 0 \nonumber \\
  \brao {J}^{\mu = 1,2,3}_3 \keto &\simeq& 0;\quad   \brao {J}^{\mu = 1,2,3}_8 \keto \simeq 0 \nonumber \\
    \brao {J}^{\mu = 1,2,3}_{QED} \keto &\simeq& 0;\quad   \brao {J}^{\mu = 1,2,3}_{Baryon} \keto \simeq 0 
 \end{eqnarray}

\end{appendices}
\end{document}